\documentclass[aps,prb,groupedaddress,showpacs,twocolumn,floatfix]{revtex4-2}
\usepackage{mathbbol,amsthm,amsmath,amsfonts,amssymb,bm,xcolor,graphicx,psfrag,array,mathtools,lipsum,mathrsfs,comment,adjustbox,stmaryrd,braket}
\usepackage{placeins}
\usepackage{natbib}

\usepackage{amsmath}
\usepackage{relsize}
\usepackage{footnote}

\usepackage{graphicx} 
\usepackage{amsmath,amssymb}
\usepackage{float}
\begin{document}
\title{Reduced probability densities of long-lived metastable states as those of distributed thermal systems:  possible  experimental implications for supercooled fluids}

\author{Zohar Nussinov}
\email{zohar@wustl.edu}
\affiliation{Rudolf Peierls Centre for Theoretical Physics, University of Oxford, Oxford OX1 3PU, United Kingdom}
\affiliation{Department of Physics, Washington University, St.
Louis, MO 63160, USA}

\date{\today}

\begin{abstract}
When liquids are cooled sufficiently rapidly below their melting temperature, they may bypass crystalization and, instead, enter a long-lived metastable supercooled state that has long been the focus of intense research. Although they exhibit strikingly different properties, both the (i) long-lived supercooled liquid state
and (ii) truly equilibrated (i.e., conventional equilibrium fluid or crystalline) phases of the same material share an identical Hamiltonian. This suggests a mapping between dynamical and other observables in these two different arenas. We formalize these notions via a simple theorem and illustrate that given a Hamiltonian defining the dynamics: (1) the {\em reduced} probability densities of all possible stationary states are linear combinations of reduced probability densities associated with thermal equilibria at different temperatures, chemical potentials, etc. (2) Excusing special cases, amongst all of these stationary states, a clustering of correlations is only consistent with conventional thermal equilibrium states (associated with a sharp distribution of the above state variables). (3) Other stationary states may be modified so as to have local correlations. These deformations concomitantly lead to metastable (yet possibly very long-lived) states. 
Since the lifetime of the supercooled state is exceptionally long relative to the natural microscopic time scales, their reduced probability densities may be close to those that we find for exact stationary states (i.e., a weighted average of equilibrium probability densities at different state variables). This form suggests several new  predictions such as the existence of dynamical heterogeneity stronger than probed for thus far and a relation between the specific heat peak and viscosities. Our theorem may further place constraints on the  putative ``ideal glass'' phase. \end{abstract}

\maketitle

\section{Introduction}

When common liquids are slowly cooled, they tend to readily crystallize to form an equilibrated solid below their freezing temperature. As known since antiquity, in the diametrically opposite limit of sufficiently rapid cooling (``supercooling'') of fluids below their freezing temperature a very different ubiquitous phase may arise. In supercooled liquids there is insufficient time for low temperature crystallization to occur \cite{Cavagna,BB}. Here, at times in excess of a material dependent minimal  waiting time $t_{\min}$ following such a supercooling of the liquid (yet below the time $t_{\sf xtal}$ required for the supercooled to overcome nucleation barriers and ultimately still crystallize (or ``devitrify'')), rich universal phenomena are seen. At these times, local few body observables and, notably, their dynamics appear to converge onto a steady state that, to this day, remains ill-understood. Numerous investigations, via careful empirical observations and simulations, have probed this state and studied its occurrence in countless settings (e.g., supercooled magma \cite{precambrian} that may form crystalline minerals at long times or, as already experimented by Fahrenheit, supercooled water that is stabilized for a long time before its crystallization into ice). This static supercooled state which appears prior to crystallization will form a focus of our attention. Here, the system empirically reaches an intriguing ``effective'' local equilibrium before achieving true global equilibrium at extremely long times $t \gtrsim  t_{\sf xtal}$ (as an equilibrium (``${\sf eq.}$'') crystalline (``${\sf xtal}$'') solid). At sufficiently low temperatures ($T \le T_{g}$), when the requisite minimal waiting time $t_{\min}$ becomes of the order of 100 seconds or larger and the system can no longer readily achieve a steady state on experimental time scales, the supercooled fluid is, in the standard taxonomy, said to be a ``glass'' \cite{Cavagna,BB}. One of the significant hurdles that conventional theories face is the disparity between the changes in the system dynamics (the relaxation times and viscosity of the supercooled fluids, especially those of ``fragile'' \cite{Angell} liquids, may increase by 15 orders of magnitude \cite{Cavagna,BB,Angell} on approaching $T_g$) with those of thermodynamic quantities and structural measures (which are, by comparison, far more modest). This disproportionate change is much unlike that conventional phase transitions. Many (often interrelated) theories, including the celebrated Random First Order Transition theory \cite{Cavagna,BB,RFOT1,RFOT2,RFOT3,RFOT4,RFOT_review} and numerous others, e.g., \cite{Cavagna, BB,Annrev17,Parisi_book,david,mode_coupling,dyn_fac,dk,tarjus,nab,dyre} have been advanced over the years. To date, nearly all theories conjecture that the supercooled state and transition into the glass are controlled by temperatures that are different from those of the equilibrium system \cite{Cavagna,BB,Steve_Gilles}. These, importantly, include an assumed transition temperature into a pristine and perfectly quiescent ``ideal glass'' \cite{Cavagna,BB,Walter,GM}. In the picture that we discuss in the current work, no special transitions are invoked. Rather, the system properties (including its excitations) can be viewed as those arising from a mixture of equilibrium solid-like and fluid-like components. 

Notwithstanding their very notable differences, {\em both the steady-state supercooled fluid and the bona-fide equilibrium system are governed by the same (disorder free) many body Hamiltonian $H$} that is defined on the full many body ($N-$particle) system $\Lambda$. Although these system may appear amorphous, the  Hamiltonian governing their dynamics is the same as that of the crystalline system and does not possess any disorder whatsoever. In this paper, our primary focus shall revolve around the examination of the probability densities associated with $n$ particles where $n$ is a small finite number that does not scale with $N$. To conform with experiment, these reduced probability densities must exhibit the said effective equilibration phenomenon of few body observables. 
If a steady state effective equilibration appears for an $n-$particle reduced probability density then it must, trivially, also appear for all reduced probability densities of particle numbers $\le n$. The attainment of local few particle steady states may take less time than those of increasing particle number (larger than $n$). As emphasized above, the true equilibration time $t_{\sf xtal}$ required for the steady state macroscopic structure
of the full many body system can be far longer. Indeed, a consequence of the theorem that we will establish is that, excusing very particular circumstances, the only steady state (for any number of particles $n\ge 2$) that satisfies locality, i.e., a  
 clustering of correlations, is none other than that of the bona fide equilibrated solid. That is, in general, a clustering of correlations precludes any other state apart from the true equilibrium state with sharp intensive state variables. Any other states (such those that we will focus on at length in the current work) are necessarily transient metastable states of finite (yet possibly still exceedingly long relative to typical microscopic time scales) lifetimes. 

 In the subsequent Sections, we will first (Section \ref{sec:stationary+near-stationary}) outline several results for general stationary states. We will then (Section \ref{sec:exp'}) proceed to discuss the possible implications of these results for experimental data with an eye towards supercooled liquids. Throughout this paper, the discussion is aimed to be non-technical so that much detail will be provided.

\section{Stationary and near-stationary reduced local probability densities}
\label{sec:stationary+near-stationary}

Expanding on Refs. \cite{1parameter,PS,longrange,critical,local_structure}, a central thesis of this paper can be succinctly summarized as follows. When accounting for all conserved quantities such as the energy $E$, average particle number, ... (and any other conserved quantities $\{ W_{\alpha}\}$ whenever these exist), the Gibbs ensemble (respectively, Generalized Gibbs ensemble when additional conserved quantities exist) stationary {\it reduced} $n-$particle probability density  
of the supercooled (or any other) system whenever it reaches a steady-state stationary state (the said ``effective'' local equilibrium) may be written as 
\begin{eqnarray}
\label{pp}
&& \rho_{{\sf stationary},n}(\beta, \mu, \cdots) = \nonumber
\\    && \int d\beta' d\mu'\ldots  P(\beta, \mu, \ldots| \beta', \mu', \ldots ) \nonumber
\\ && \times ~~\rho_{{\sf eq.},n}(\beta',\mu', \ldots).
\end{eqnarray}
Eq. (\ref{pp}) relates the reduced $n-$particle probability density ($\rho_{{\sf stationary},n}$) of the  stationary system to that of the equilibrium system ($\rho_{{\sf eq.},n}$). Classically, these two reduced probability densities are functions of the $n-$particle phase space variables. To streamline notation, we have omitted in Eq. (\ref{pp}) the explicit inclusion of $n$-particle phase space coordinates (or other particle labels) as arguments for the reduced probability densities. Quantum mechanically, these reduced probability densities are functions of a set of mutually compatible coordinates or quantum numbers of the $n-$particle subsystem. We will often drop the explicit ``${\sf stationary}$'' subscript in Eq. (\ref{pp}). Unless stated otherwise, reduced probability densities will refer to those of the stationary system. In Eq. (\ref{pp}), we highlight state variables such as the inverse temperature $\beta$ (or $\beta'$) that parameterize the supercooled (respectively, equilibrium) systems. In the ellipsis of Eq. (\ref{pp}), additional dependencies on all constants of motion $\{\mathcal{W}_{\alpha}\}$ may appear. As will become apparent, under rather modest conditions, Eq. (\ref{pp}) applies to all stationary states, regardless of their number or specific character in a given system.

In Ref. \cite{longrange}, it was illustrated that in driven systems, the energy density can display finite standard deviation fluctuations (as measured by the full system probability density $\rho_{\Lambda}$) and that general correlations need not be local. In such systems \cite{1parameter,PS,longrange,critical,local_structure},
an analogue of Eq. (\ref{pp}) holds for the full many body probability density. It is natural to assume that upon the cessation of external driving forces, local self-generated noise that is uncorrelated across the system may destroy long-range correlations and effectively equilibrate the system. Nonetheless,  variations in local quantities can still persist. Locally, Eq. (\ref{pp}) may still hold with a nontrivial kernel $P$ as we will examine in some depth. Indeed, our focus in the current work will be on local reduced few particle probability densities.

The reduced equilibrium probability density $\rho_{{\sf eq.},n}$ is completely static as the system evolves under its own Hamiltonian. Thus, it is clear that, whenever it is realized, any candidate probability distribution $\rho_{{\sf stationary},n}(\beta, \mu, \cdots)$ that is of the form of Eq. (\ref{pp}) with a time independent $P(\beta, \mu, \ldots| \beta', \mu', \ldots )$ must be stationary. The relation of Eq. (\ref{pp}) is far stronger and constitutes an ``if and only if'' condition for stationarity. Indeed, as we will prove, under modest assumptions,  \bigskip

$\bullet$ {\it all stationary reduced probability densities must be of the form of Eq. (\ref{pp}).}  \bigskip

A key point is that the expectation values of local observables in the nearly stationary state of the supercooled fluid and other systems are different from those in the true equilibrium state. This implies that if one may treat the supercooled liquid as truly stationary then we will further have that  \bigskip

$\bullet$ {\it the function $P$ in Eq. (\ref{pp}) cannot be the product of delta functions}
\begin{eqnarray}
\label{different_is_more}
P(\beta, \mu, \ldots| \beta', \mu', \ldots ) \neq \delta (\beta' - \beta) \delta(\mu' - \mu) \cdots .
\end{eqnarray}
The same holds true for {\it any static metastable state}. \\  

We explicitly note that apart from the conventional state variable such as the inverse temperature and chemical potential, pressure, etc., the  variables appearing in Eqs. (\ref{pp}, \ref{different_is_more}) may, e.g.,  also include external fields or externally imposed stresses on the system (or their conjugate variables such as the magnetization, polarization, or elastic strains).

The remainder of this Section is organized as follows. In Section \ref{sect:proof_eqpp}, we will first consider the general completely stationary in time $n-$particle states without imposing spatial locality and establish Eq. (\ref{pp}) as a theorem. We will then discuss the exceedingly slow empirical temporal variations of the reduced particle densities of the supercooled fluid in Section \ref{finite_time} and thereafter turn to general discussions of spatial locality of the reduced probability densities in Section \ref{locality} and Appendix \ref{nearly-commute}. In Section \ref{sub:hydro}, we will relate our considerations to hydrodynamic systems.  To make our more general arguments clear, we will regress along the way (Sections \ref{decoupled_solvable} and \ref{1d-decoupled_particles}) to one dimensional and other simple example models. These models are rather special cases rather than the rule yet they shed additional light on the structure of our theory. In Section \ref{consequences}, we will outline the general predictions that follow from our framework of Eq. (\ref{pp}). Lastly, in Section \ref{sec:eff}, we will introduce the notion of ``effective'' many body probability densities that may yield the reduced probability densities of Eq. (\ref{pp}) (and thus may be used to calculate few body expectation values) yet do not necessarily describe the many body system. 

\subsection{A central relation- constraints on reduced static probability densities}
\label{sect:proof_eqpp}
We now formalize our discussion via a simple theorem. \\

{\bf{Theorem.}} So long as equilibrium ensemble equivalence holds for a given Hamiltonian $H$ and the state variables $\beta,' \mu', \cdots$ completely specify the reduced equilibrium probability density then {\it any such reduced probability density} that is stationary under evolution with $H$ can be written in the form of Eq. (\ref{pp}) with an appropriate conditional probability $P$.  \\

{\it Proof.} By definition, the reduced $n-$body probability density can be written as the partial trace of the full many body probability density,
\begin{eqnarray}
\label{full}
    \rho_{n} \equiv {\sf Tr}_{n+1, \ldots, N} ~\rho_{\Lambda}.
\end{eqnarray}
Classically, the above partial trace is replaced, in the usual manner, by integrals over the phase space coordinates of particles $n+1, \ldots, N$ followed by an explicit symmetrized average (over subgroups of identical particles out of the $N$ particles in the system). From Eq. (\ref{full}), each component $(\tilde{\rho}_n(\omega) e^{-i \omega t})$ of the probability density associated with oscillations of  frequency $\omega$ satisfies
\begin{eqnarray}
\label{trivw}
    \tilde{\rho}_n(\omega) = {\sf Tr}_{n+1, \ldots, N}~ \tilde{\rho}_{\Lambda}(\omega).
\end{eqnarray}
In particular, any static (i.e., $\omega=0$) component of the reduced probability density \footnote{This includes, e.g., for the case of the supercooled liquid, any numerically obtained/experimentally measured long time average of the reduced probability density in the time interval $t_{\sf xtal} \ge t  \ge t_{\min}$.} 
can be expressed as a partial trace of a corresponding static many body density matrix $\rho_{\Lambda}$  \footnote{For the example of the supercooled liquid, this refers to the long time average of $\rho_{\Lambda}$ within the time interval $t_{\sf xtal} \ge t \ge t_{\min}$ where the average few body observables are nearly static and the system appears to reach an effective equilibrium.}. By the von Neumann equation $\partial_{t} \rho_{\Lambda} = \frac{i}{\hbar}[\rho_{\Lambda}, H]$ 
(or its classical Liouville equation counterpart $\partial_t \rho_{\Lambda} = \{ H, \rho_{\Lambda}\}_{P.B.}$ with ``$P.B.$'' denoting the Poisson bracket), any stationary component of the many body density matrix must trivially commute (or have a vanishing Poisson bracket) with the system Hamiltonian. 
When discussing the zero-frequency component, we allude here to the long time average of the reduced probability density in the limit of large averaging times $\tau$,
\begin{eqnarray}
\label{rlta}
\rho_{n,l.t.a}(\tau) \equiv \frac{1}{\tau} \int_{0}^{\tau} \rho_{n}(t')~ dt'.
\end{eqnarray}
If the reduced probability probability density $\rho_{n}(t')$ is stationary (the condition of the Theorem) then it will, of course, be equal to its long time ($\tau \to \infty$) average.
(In later comparisons to empirical data, our focus will be on the time window $\tau$ where the reduced probability density is nearly stationary. For a large yet finite $\tau$, there may  be additional ${\cal{O}}(1/\tau)$ corrections as will be briefly discussed in Section \ref{finite_time}.) We introduce an analogous definition for the long time average $\rho_{\Lambda, l.t.a.}(\tau)$ of the probability density $\rho_{\Lambda}$ of the full system, i.e.,
\begin{eqnarray}
\label{llta}
\rho_{\Lambda,l.t.a}(\tau) \equiv \frac{1}{\tau} \int_{0}^{\tau} \rho_{\Lambda}(t')~ dt'.     
\end{eqnarray}
 Given Eq. (\ref{llta}), for any finite $\tau$, an $\omega =0$ counterpart of Eq. (\ref{trivw}) follows by averaging both sides of Eq. (\ref{full}) over the time interval $[0,\tau]$ \footnote{Since at any time $t'$, the full many body probability density $\rho_{\Lambda}(t')$ is normalized, it satisfies Fubini's theorem. In other words, the order of the partial trace in Eq. (\ref{full}) and the integration over the time variable such as that associated with the averaged long time integral of Eq. (\ref{rlta}) can be interchanged. This interchange leads to Eq. (\ref{ft}) with the definition of  Eq. (\ref{llta}) (and analogously leads to Eq. (\ref{trivw}) when noting that the averaged time integral of the full trace of $|\rho_{\Lambda}(t') e^{i \omega t'}|$ is also trivially normalized- thus similarly satisfying the condition for Fubini's theorem).},  
\begin{eqnarray}
\label{ft}
    \rho_{n,l.t.a}(\tau) = {\sf Tr}_{n+1, \ldots, N}~ \rho_{\Lambda, l.t.a.}(\tau).
    \end{eqnarray}
     In a $\tau \to \infty$ limit when all time dependence is integrated over, the long-time average of the many body probability density $\rho_{\Lambda}$ is, by construction, a time independent static quantity. 
    
We will next investigate what are the most general possible static $\rho_{\Lambda, l.t.a.}$. This will then allow us to determine all possible time independent reduced probability densities $\rho_{n,l.t.a}$  and illustrate that these must be of the form of Eq. (\ref{pp}). We will explicitly separately consider 
 (a) {\it quantum}  and (b) {\it classical} theories
 and then illustrate how in both of these,
 a statement about the long-time average of the many body probability density $\rho_{\Lambda}$ 
(Eq. (\ref{longpp+})) leads to Eq. (\ref{pp}).  \\

(a) For {\it quantum systems,} it follows from the above noted stationarity in the $\tau \to \infty$ limit that 
\begin{eqnarray}
\label{comm}
    [\rho_{\Lambda,l.t.a.},H]=0.   
\end{eqnarray}
Equivalently, since the eigenvalues of any probability density matrix including those of $\rho_{\Lambda}(t')$ are positive semidefinite and sum to unity, the largest norm amongst those of all eigenvalues, i.e., the operator norm of the probability density at general times $t'$, 
\begin{equation}
\label{boundedrho}
  ||\rho_{\Lambda}(t') || \le 1.  
\end{equation}
Thus, Eqs. (\ref{llta}, \ref{boundedrho}) and the von-Neumann equation imply that the operator norm  
\begin{eqnarray}
\label{long+triv+}
&& \Big| \Big|[\rho_{\Lambda,l.t.a.},H] \Big| \Big|  =
    \Big| \Big|\frac{1}{\tau } \int_{0}^{\tau}
    dt' [\rho_{\Lambda}(t'),H] \Big| \Big| \nonumber
    \\ && = \frac{\hbar}{\tau} \Big| \Big |\int_{0}^{\tau} dt' \frac{d \rho_{\Lambda}(t')}{dt'} \Big| \Big|  =   \frac{\hbar}{\tau} \Big| \Big |\rho_{\Lambda}(\tau) - \rho_{\Lambda}(0) \Big| \Big| \nonumber
    \\ && \le  \frac{2\hbar}{\tau}.
\end{eqnarray}
In the $\tau \to \infty$ limit, the upper bound in the last line of Eq. (\ref{long+triv+}) vanishes thus  reestablishing Eq. (\ref{comm}). 
This vanishing commutator implies that all static reduced probability densities (equal to the long time average $\rho_{n,l.t.a.}$ of Eq. (\ref{rlta}) in the $\tau \to \infty$ limit) can be expressed as a partial trace, 
    \begin{eqnarray}
    \label{ntrLambdalta}
    \rho_{n,l.t.a} = {\sf Tr}_{n+1, \ldots, N} ~~\rho_{\Lambda, l.t.a.},
    \end{eqnarray}
    of any many body probability density $\rho_{\Lambda, l.t.a.}$ that may be simultaneously diagonalized with the Hamiltonian $H$. We now write the most general operators that may commute with $H$ (i.e., the most general solutions to the  stationary condition of Eq. (\ref{comm})). In what follows, $\{W_{\alpha}\}$ mark the eigenvalues of all integrals of motion that commute with the Hamiltonian (energy eigenvalues $E$) with $|E, \{W_{\alpha}\}\rangle$ the corresponding mutual eigenstates. Given Eq. (\ref{comm}), the operators
    \begin{eqnarray}
    \label{Pproj}
        {\mathbb{P}}_{E, \{W_{\alpha}\}} \equiv |E, \{W_{\alpha}\}\rangle \langle E, \{W_{\alpha}\} |
        \end{eqnarray}
        form a complete set of orthogonal projections (with respect to the (Hilbert-Schmidt) trace inner product) that span the space of stationary states $\rho_{\Lambda}$ over the full $N-$body Hilbert space. Any individual projection operator ${\mathbb{P}}_{E, \{W_{\alpha}\}}$ is associated with the many body microcanonical ensemble probability density matrix (in the absence of quantum numbers $\{W_{\alpha}\}$ other than the system energy) or its generalized variant (when there are additional quantum numbers $\{W_{\alpha}\}$). Thus, the most general probability density satisfying Eq. (\ref{comm}) is a normalized linear combination of the projection operators of Eq. (\ref{Pproj}). \\
        
       (b) {\it For classical systems},    the stationarity of $\rho_{\Lambda, l.t.a.}$ in the $\tau \to \infty$ limit similarly implies that $\rho_{\Lambda, l.t.a.}$ must be expressible as a function of the energy and (if any exist) all other constants of motion. Here, the commutator of Eq. (\ref{comm}) is trivially  replaced by a Poisson bracket,
        \begin{eqnarray}
\label{comm'}
    \{\rho_{\Lambda,l.t.a.},H\}_{P.B.}=0. 
\end{eqnarray}
Alternatively, one may note that since the probability density is positive semidefinite, $\rho_{\Lambda}(t') \ge 0$, and integrates to a value of one over all of phase space it follows that at any time $t$, the integral of the norm $|\rho_{\Lambda}(t')|$ over all of phase space is trivially equal to one. Thus, from the Liouville equation and the definition of the long time average of Eq. (\ref{llta}), it is readily seen that the integral of the absolute value of the lefthand side of Eq. (\ref{comm'}), i.e., the integral of $|\{\rho_{\Lambda,l.t.a.},H\}_{P.B.}|  =  \frac{1}{\tau}|\rho_{\Lambda}(\tau) - \rho_{\Lambda}(0)|$, over all of phase space is bounded from above by $\frac{2}{\tau}$ (an upper bound which tends to zero in the $\tau \to \infty$ limit and hence establishes the equality of Eq. (\ref{comm'})). The operators $\{{\mathbb{P}}_{E, \{W_{\alpha}\}}\}$ of Eq. (\ref{Pproj}) are now replaced by projections to phase space surfaces of constant energy $E$ 
        (i.e., microcanonical ensemble phase space shells of energy $E \le H(
x,p) \le E + \Delta E$) and the conserved quantities $\{{\mathcal{W}}_{\alpha}(x,p)\}$ with $(x,p)$ denoting the $N-$particle phase space coordinates
        \footnote{As is well known, e.g. \cite{LL1}, for a system with $f$ degrees of freedom, there may be $(2f-1)$ conserved quantities when integrating the corresponding equations of motion (with a second order differential equation (and thus, apart from a trivial time  shift, two constants of motion) for each degrees of freedom); our focus is on the symmetry borne additive conserved quantities that lead to intensive state variables in the thermodynamic limit.}.  
        These projections will then be weighted with a normalized probability distribution so as generate the most general static probability density $\rho_{\Lambda,l.t.a.}$ that may depend only on the constants of motion of the system. The phase space projection operators explicitly read 
        \begin{eqnarray}
        \label{classicalP}
        && {\mathbb{P}}_{E, \{W_{\alpha}\}} = \Theta(E+ \Delta E - H(x,p)) \nonumber
        \\ && \times   \Theta(H(x,p)-E) \prod_\alpha \Big( \Theta(W_\alpha+ \Delta W_\alpha-\mathcal{W}_\alpha(x,p)) \nonumber
        \\ && \times \Theta(\mathcal{W}_{\alpha}(x,p) - W_\alpha) \Big).
        \end{eqnarray}
        The values of the constants $\Delta E, \{\Delta W_{\alpha}\}$ in Eq. (\ref{classicalP}) set the system size independent finite 
 width of the fluctuations of the Hamiltonian (and the other conserved quantities $\mathcal{W}_{\alpha}$) about the energy $E$ (and expectation values $W_\alpha$). That is, the effective delta function product $(\delta(H-E) \prod_{\alpha} \delta(\mathcal{W}_{\alpha} - W_{\alpha}))$ of Eq. (\ref{Pproj}) is replaced by a narrow shell allowing for finite deviations of $H$ and $\mathcal{W}_\alpha$ when these are continuous quantities. For discrete $\mathcal{W}_{\alpha}$, the Heaviside $\Theta$-function products in Eq. (\ref{classicalP}) are replaced by Kronecker deltas. The inner trace product between operators in the quantum arena is replaced by a phase space overlap integral. \\ 
        
        We next return to general considerations that apply to both quantum and classical systems. Invoking Eq. (\ref{comm}) for quantum systems or Eq. (\ref{comm'}) for their classical counterparts, it is seen that any stationary many body probability density is a normalized (${\sf Tr}_{\Lambda} ~\rho_{\Lambda} =1$) weighted linear combination of these projection operators,
        \begin{eqnarray} 
        \label{longpp+}
            \rho_{\Lambda, l.t.a} \ = \int dE'  \int \prod_{\alpha} dW'_{\alpha}  {\cal P}(E', \{W'_{\alpha}\}) 
            {\mathbb{P}}_{E', \{W'_{\alpha}\}} \nonumber
            \\ = \int dE'  \int \prod_{\alpha} dW'_{\alpha}  {\cal P}(E', \{W'_{\alpha}\}) 
            \rho^{\sf m.c.}_{\sf \Lambda, eq., E', \{W'_{\alpha}\}}.
        \end{eqnarray}
        Here, ${\cal{P}}(E', \{W'_{\alpha}\})$ denotes the probability for having an energy $E'$ and quantum numbers (conserved classical quantities) $\{W'_{\alpha}\}$ in the long time average of the probability density $\rho_{\Lambda, l.t.a}$. In Eq. (\ref{longpp+}), $\rho^{\sf m.c.}_{\sf \Lambda, eq., E', \{W'_{\alpha}\}}$ is the probability density for the equilibrium microcanonical (or Generalized microcanonical ensemble) for a the system $\Lambda$ given an energy $E'$ (and other conserved numbers $ \{W'_{\alpha}\}$ when present). For smooth ${\cal P}$,
        the domain of integration may be split into small intervals over the energy and $W'_{\alpha}$ values and for each such interval, the integral of $( {\cal P}(E',\{W'_{\alpha}\}) {\mathbb{P}}_{E', \{W'_{\alpha}\}})$ may be replaced by 
        ${\cal P}$ acted on by a classical phase space or similar quantum projections to $N-$ body states that lie in {\it a finite width} energy window about a fixed energy 
        $E'$ (and, if applicable, a similar projection to other conserved numbers $\{W'_{\alpha}\}$). The latter sum may then be replaced by a microcanonical (or, respectively, further constrained microcanonical ensemble) average value \footnote{If the function ${\cal P}$ is not smooth then Eq. (\ref{longpp+}) will be valid only if an extreme form of general ensemble equivalence (a condition for our Theorem)- the Eigenstate Thermalization Hypothesis (ETH)
 \cite{ETH1,ETH2}- holds. The ETH asserts that one can take the microcanonical ensemble in the limiting form of a single energy eigenstate. Whenever the ETH holds, the projection operator
 to each individual eigenstate inasmuch as local observables are concerned may be replaced by an equilibrium probability density set by the state variables $E', \{W'_{\alpha}\}$. In systems in which the ETH is not satisfied (e.g., those with many body localization \cite{MBL,MBL_rev} or in scar \cite{scar,scar_rev} states), Eq. (\ref{longpp+}) need not be satisfied.}. Eq. (\ref{longpp+}) implies that the reduced $n-$particle probability density of the stationary system,
        \begin{eqnarray}
        \label{longpp}
           \rho_{{\sf{stationary}},n}(\beta,\mu, \cdots) =  {\sf Tr}_{n+1, \cdots, N}~ \rho_{\Lambda, l.t.a} \nonumber
           \\  =  \int dE'   \int \prod_{\alpha} dW'_{\alpha} ~  {\cal P}(E', \{W'_{\alpha}\}) \nonumber
            \\ \times {\sf Tr}_{n+1, \cdots, N} ~\Big( \rho^{\sf m.c.}_{\sf \Lambda, eq., E', \{W'_{\alpha}\}}\Big) \nonumber
            \\ \equiv \int d \beta' d \mu' \ldots ~ P (\beta, \mu, \ldots| \beta', \mu', \ldots) \nonumber
            \\ \times \rho_{{\sf eq.},n}(\beta', \mu', \cdots).
        \end{eqnarray}
         In the last equality of Eq. (\ref{longpp}), we invoked the condition of assumed ensemble equivalence, 
        \begin{eqnarray}
        \label{ensemble_equiv}
        \rho_{{\sf{eq.}},n}(E', \cdots) \equiv {\sf Tr}_{n+1, \cdots, N} ~\Big( \rho^{\sf m.c.}_{\sf \Lambda, eq., E', \{W'_{\alpha}\}}\Big) \nonumber
        \\ = \rho_{{\sf{eq.}}, n}(\beta', \mu', \cdots),
        \end{eqnarray}
        and accordingly set, for the fixed values of $\beta, \mu, \cdots$,
        \begin{eqnarray}
        \label{cond_prob}
        P (\beta, \mu, \ldots| \beta', \mu', \ldots) \equiv {\cal P}(E',N', \{W'_{\alpha}\}),
        \end{eqnarray}
        where $N'$ is the number of particles of the equilibrium system with the conserved quantities 
        $\{W'_{\alpha}\}$ replaced by intensive variables in the ellipsis in the argument of the $P$ \footnote{Observe that Eq. (\ref{cond_prob}) makes no assumptions about the state of the stationary system. The function ${\cal P}(E',N', \{W'_{\alpha}\})$ may be a rather rapidly varying one. 
        All that is important is that the equilibrium energy $E'$ is associated with a unique inverse temperature $\beta'$
        for which $E' = U(\beta')$ with $U$ the internal energy of the equilibrium system and that, similarly, the particle number $N'$ is uniquely determined by the chemical potential $\mu'$. In Eq. (\ref{pp'}), we will discuss situations (e.g., those of equilibrium phase coexistence) in which an extensive range of equilibrium energies $E'$ (finite range of equilibrium energy densities) is associated with a single inverse temperature $\beta'$.}. 
        Given Eq. (\ref{cond_prob}), we may, for any set of fixed state variables $\beta, \mu, \ldots$ of the stationary system, view $P$ as the conditional probability of having equilibrium state variables $\beta', \mu', \ldots$. Putting all of the above pieces together, Eqs. (\ref{longpp}-\ref{cond_prob}) establish Eq. (\ref{pp}) for {\it stationary} 
        reduced probability densities when the $\tau \to \infty$ limit may be considered. $\blacksquare$ \\

        A consequence of our proof is that for the full system $\Lambda$ (i.e., formally replacing $n$ in the above expressions by the total particle number $N$), a stationary probability density must similarly satisfy Eq. (\ref{pp}). This corollary is indeed apparent from our intermediate step of Eq. (\ref{longpp+}) and was explored earlier  \cite{1parameter,PS,longrange,critical,local_structure} when focusing on global probability density (not the reduced probability density that we examine in the present work) of stationary systems. Our current  interest centers on system averaged few particle observables. As noted in the Introduction, these few body expectation values can appear to converge onto nearly stationary values in the supercooled state (while the entire many body fluid is not stationary \footnote{For the supercooled liquid, the non-stationarity of the full many body system is, e.g., evinced by the time dependent character of the spatially heterogeneous dynamics of the liquid \cite{Cavagna,BB,DH2,DH3,rotation}).}). In such instances, corrections to the general limit of completely stationary few body reduced probability density of Eq. (\ref{pp}) might be anticipated to be small, see Section \ref{finite_time} (whereas dynamical corrections to the full probability density of the system may be more apparent since only the measured few body expectation values appear to converge onto a nearly steady state). As we will illustrate in substantial detail in some of the forthcoming Sections, excusing systems that evolve trivially in time (similar to the examples to be outlined in Sections \ref{decoupled_solvable} and \ref{1d-decoupled_particles}), generally, if a clustering of correlations is satisfied (i.e., if the covariance between local observables decays with distance) then the only possible conditional probability kernel $P$ is a delta function in the equilibrium state variables. That is, not too surprisingly, the only stationary state of the full $N-$body system is the  equilibrium configuration (having sharp unique values of all of its intensive state variables).

In Section \ref{sec:eff}, we will discuss exactly stationary states of the full many body system that generally do not satisfy a clustering of correlations (i.e., a decay of connected correlations at large distances). Nonetheless, the latter exactly stationary states will exhibit expectation values of few ($n$) particle local observables that are identical to the expectation values of the same observables when these are evaluated with nearly stationary probability densities (where these  probability densities will exhibit a clustering of correlations). 

The probability distribution of Eq. (\ref{different_is_more}) may describe metastable non-equilibrium (``{\sf neq}'') systems other than supercooled fluids that we will focus on in this paper when discussing experimental consequences (Section \ref{sec:exp'}). For instance, for Ising ferromagnets undergoing hysteresis in an external magnetic field, the relevant conditional probability satisfies 
Eq. (\ref{different_is_more}) which we rewrite here with an explicit magnetization dependence,
\begin{eqnarray}
\label{pneqferro}
 && P_{\sf neq, ferro.}(\beta,\mu, M \ldots|\beta',\mu', M' \ldots) \nonumber
 \\ && \neq \delta (\beta'- \beta) \delta(\mu' - \mu) \delta (M'-M) \cdots.
 \end{eqnarray}
In Eq. (\ref{pneqferro}), $M$ and $M'$ denote, respectively, the magnetization of the equilibrium system and the non-equilibrium system in the given external magnetic field $h$. Such a ferromagnet in an external magnetic field may appear to be effectively  ``stuck'' in an equilibrium state of another Hamiltonian (one associated with a  different, earlier applied, external field $h''$ on the ferromagnet along the hysteresis loop). The pertinent conditional probability in 
Eq. (\ref{pneqferro}) might still be a of a canonical delta-function form in $M'$,
\begin{eqnarray}
\label{npferrom}
&& P_{\sf neq, ferro.}(\beta,\mu, M \ldots|\beta',\mu', M' \ldots) \nonumber
\\ && \propto  \delta (M'-M''),
\end{eqnarray} 
yet, importantly, with the latter sharp magnetization $M'' \neq M$. Here, $M''$ is the magnetization of an equilibrium system placed in an external applied field $h''\neq h$. We reiterate that this magnetization $M''$ is indeed different from the expected equilibrium magnetization $M'$ when the applied external magnetic field is equal to $h$ (the one in which the system is actually in). Since, apart from the shift in the magnetization (and other similar state variables) the distribution of Eq. (\ref{npferrom}) is of a  canonical equilibrium form, it can be associated with a free energy. Such a free energy is metastable relative to the true free energy of an equilibrium system of   magnetization $M$.

By contrast, in the case of the supercooled fluid that we will examine in Section \ref{sec:exp'}, empirical observations dictate that the relevant state variables $\beta', \mu', \ldots$ {\it cannot} be set to fixed values to describe the equilibrium (either liquid or crystalline) system at different state variables. For supercooled fluids, the probability distribution $P$ of Eq. (\ref{pp}) cannot be a simple product of delta functions.

        In general, at the transition points/lines of the equilibrium system, the probability 
        density $\rho_{\sf eq}(\beta',\mu', \ldots)$ may be ill-defined.
In these situations, $\beta', \mu', \ldots$ do not  completely specify the equilibrium state. At a typical discontinuous transition with coexisting phases, a range of energy densities or number densities may, e.g., be associated with a single inverse temperature $\beta'$ or chemical potential  $\mu'$ of the equilibrium system. We denote such a range of energy densities as the ``Phase Transition Energy Interval'' (${\cal PTEI}$). In these situations, in order to satisfy the conditions underlying our Theorem (i.e., to have state variables that uniquely specify the reduced equilibrium probability density), we need to amend Eq. (\ref{pp}) by a more comprehensive form. That is, we must account for the possible variation of the equilibrium energy density $(\epsilon')$, particle number $({\tilde{{\sl n}}}')$, and other densities at fixed temperature, chemical potential, or other fixed state variables. In what follows, we denote by ${\tilde{{\sl n}}}$ the particle number density in the stationary state. We may  replace the integration over the energies in Eq. (\ref{longpp+}) by an integration over the equilibrium values of $\beta'$ only for those intervals of the equilibrium energy density $\epsilon'$, equilibrium number density ${\tilde{\sl n}}'$, and other intensive  state variable densities that do not lie in the ${\cal PTEI}$. This yields 
\begin{eqnarray}
\label{pp'}
&& \rho_{{\sf stationary},n}(\beta, \mu, \cdots) = \nonumber
\\    && \int d\beta' d \mu' \ldots \Big( P(\beta,  \mu, \ldots| \beta', \mu', \ldots ) \nonumber
\\ && \times ~~\rho_{{\sf eq.},n}(\beta',  \mu', \ldots) \Big) \nonumber
\\ && + \int_{\cal{PTEI}} d\epsilon' d {\tilde{{\sl n}}}'  \cdots 
\Big(P(\epsilon, {\tilde{{\sl n}}},  \ldots| \epsilon', {\tilde{{\sl n}}}'
,\ldots) \nonumber
\\ && \times \rho_{{\sf eq.},n} (\epsilon', {\tilde{{\sl n}}}',  \ldots) \Big). 
\end{eqnarray}
Eq. (\ref{pp'}) expresses stationary states as weighted sums of equilibrium probability distributions. One may readily extend this expression to one involving different varying composition by explicitly including chemical potentials $\{\mu'\}_{a=1}^{q}$ and particle number densities $\{{\tilde{{\sl n}}}_{a}'\}_{a=1}^{q}$, etc., for multi-component ($q>1$) systems. For simplicity, we dispense with an explicit inclusion of these or other independent state variables (e.g., pressures) in Eq. (\ref{pp'}). As is well known, phase coexistence and metastability are often analyzed in mean-field type theories by free energy considerations and Maxwell constructs. Such mean-field type effective free energies are designed to capture quintessential features of the different phases and their local  interfaces. To avoid confusion, we underscore that in the second integral of Eq. (\ref{pp'}), $\rho_{{\sf eq.},n}(\epsilon', {\tilde{{\sl n}}}',  \ldots)$ is already the exact reduced body probability density of the system within the equilibrium mixed phase coexistence (${\cal{PTEI}}$) regime.

 \subsection{Solvable limit of decoupled subsystems}
   \label{decoupled_solvable}

As an example system for which there is a nontrivial distribution $P$ in Eq. (\ref{pp}) (i.e., one with a spread of $\beta'$ values), 
we consider a system of distinguishable decoupled particles (either classical or quantum) that may realize different single particle states. Here, 
    \begin{eqnarray}
    \label{trivexample}
        H &=& \sum_{i=1}^{N} H_{i}, \nonumber
        \\ \rho_{\Lambda}(\beta) &=& \prod_{i=1}^{N} \rho_{1,i}(\beta)
    \end{eqnarray}
    In Eq. (\ref{trivexample}), $\{H_i\}$ are the Hamiltonians of the different individual decoupled particles
    and $\rho_{1,i}$ denotes the reduced single particle probability density for particle $i$. We further consider the  situation in which, uniformly for each particle $i$, 
    \begin{eqnarray}
    \label{eq:rho1}
\rho_{1,i}(\beta) = \rho_{1}(\beta)  = \int d\beta'~ P(\beta|\beta') ~\rho_{{\sf eq},1}(\beta').
\end{eqnarray}
By construction, Eq. (\ref{pp}) is satisfied here exactly. Similarly, for any inter-spin spatial separation, the reduced two-spin probability density is equal to the product of the reduced single particle probability densities. Thus, the covariance between single site operators vanishes identically, 
\begin{eqnarray}
\label{ABfactor}
 {\sf Tr}(\rho_{2}(\beta)AB) - ({\sf Tr}(\rho_1 (\beta) A))({\sf Tr}(\rho_{1}(\beta)B))=0.
 \end{eqnarray}
 Given the decoupled nature of the Hamiltonian (which is a sum of such single site operators) and the state of the system, the energy, 
\begin{eqnarray}
\label{Hscaling}
E= \langle H \rangle = N ~ {\sf Tr}(\rho_{1} H_{1}), 
\end{eqnarray}
with a standard deviation
\begin{eqnarray}
\label{sigmaHscaling}
\sigma_{E} = \sqrt{N} ~ \sigma_{H_1}. 
\end{eqnarray}
That is, even though the standard deviation of the energy density $H/N$ tends to zero (as $N^{-1/2}$) in the thermodynamic (large $N$) limit, there is an effective distribution of the temperatures in the reduced single particle probability density. As we will later explain (Eq. (\ref{non_equil_long})), if Eq. (\ref{pp}) holds for reduced two-particle probability densities for arbitrary far separated sites then unless $P(\beta|\beta')$ is a delta-function in $\beta'$, the covariance between distant particles (spins) will not vanish (see also \cite{longrange}). In the current example, Eq. (\ref{pp}) holds only strictly locally- i.e., for single ($n=1$) particle (spin) probability distribution alone. 
The two-particle (spin) probability distribution derived from the many body $\rho_{\Lambda}$ of Eq. (\ref{trivexample}) trivially factorizes into single particle probability densities, $\rho_{2}(i,j) = \rho_{1}(i) \rho_{1}(j)$, regardless of the distance between the particles (spins). In more generic situations, when finite distance corrections may appear for the latter factorization (i.e., connected correlations are present between particles (or spins) that are a finite distance away) yet $\rho_{2}(i,j) \to \rho_{1}(i) \rho_{1}(j)$ as the distance between the two particles (or spins) diverges; Eq. (\ref{ABfactor}) will only be valid for asymptotically large distances. If $|\rho_{2}(i,j) - \rho_{1}(i) \rho_{1}(j)|$ is bounded from above by a power law function in the distance between particles (spins) $i$ and $j$  then so will the covariance on the lefthand side of Eq. (\ref{ABfactor}) for general local operators $A$ and $B$ and the scaling relations of Eqs. (\ref{Hscaling}, \ref{sigmaHscaling}) for $(E/N)= {\cal{O}}(1)$ and $(\sigma_{E}/N) = {\cal{O}}(N^{-1/2})$ will remain unchanged in the thermodynamic limit.

The local Hamiltonians and probability densities in Eqs. (\ref{trivexample}) could, e.g., correspond to non-interacting spins in an external magnetic field, i.e., the single spin Hamiltonian  
\begin{eqnarray}
\label{sitefield}
    H_{i,{\sf field}} = - h \sigma_i^z
\end{eqnarray} 
and its associated equilibrium probability density or, e.g., a classical Ising spin-chain in an external (longitudinal) field. In the latter case, setting \begin{eqnarray}
\label{siteIsing}
   H_{i,{\sf Ising}}= -\sigma^z_i 
 \sigma^z_{i+1}  
\end{eqnarray}
in $H_{{\sf Ising}} = \sum_{i=1}^{N-1} H_{i,{\sf Ising}}$ leads, up to a additive constant and a boundary term, to the  Hamiltonian of the nearest neighbor Ising chain.
  
For either of the Hamiltonians of Eqs. (\ref{sitefield}, \ref{siteIsing}), the decoupled $\{H_i\}$ can be regarded as single site variables.
For the nearest neighbor Ising interaction of Eq. (\ref{siteIsing}) this can be done following a simple  duality transformation mapping $H_{i,{\sf Ising}}$ to a single Ising spin and writing the equilibrium probability densities $\rho_{1,eq}$ and ensuing system probability density $\rho_{\Lambda}$ of Eq. (\ref{trivexample}) for these. Thus, we see that simple interacting systems like a classical Ising chain in a field are captured by the example of Eq. (\ref{trivexample}). In such classical examples, all of the Hamiltonians $H_i$ (and the individual spins $\sigma^z_{i}$)  somewhat tautologically commute with one another. Here, one may define an extensive number of conserved quantities $\{\beta_i\}_{i=1}^{N}$ associated with these local Hamiltonians (or, equivalently, with each of the $N$ individual spins $\sigma^z_{i}$). 

The above considerations generalize trivially to Hamiltonians that are separable and states that may be expressed as a product of  
$n-$particle thermal states
as in the second of Eqs. (\ref{trivexample}) yet now involving $n$ particles with general finite $n\ge 1$,
\begin{eqnarray}
\label{divr}
    \rho_{\Lambda} = \prod_{a = 1}^{N/n} \rho_{n}(a). 
\end{eqnarray}
In Eq. (\ref{divr}), the probability density of the full $N$ particle system factorizes into $(N/n)$ decoupled (reduced) probability densities for each of the $n$ particle subsystems. The procedure that we have undertaken above (keeping clustering in tact as evinced in  Eq. (\ref{divr})) 
may be repeated verbatim here.

To gain some further intuition by comparison to trivial limiting case, we may consider the reduced $s-$body 
(with $n$ being an integer multiple of $n'$) probability density
\begin{eqnarray}
\label{rhoc}
  \rho^{c}_{n} = \prod_{i=1}^{n/n'} \rho_{{\sf eq},n'}(\beta'_i,i) 
\end{eqnarray}
(similar to that considered in Eq. (\ref{divr}) yet now the product of the reduced $n'-$particle {\it equilibrium} probability densities at inverse temperatures $\beta'_i$ constituting a reduced few ($n-$)particle probability densities). Since it is a product of equilibrium probability densities, $\rho^{c}_{n}$ is stationary and is thus, according to our Theorem, expressible in the form of Eq. (\ref{pp}). Indeed, for distinguishable particles, $\rho^{c}_{n}$ can, e.g., be written as the partial trace
of the stationary full $N-$particle system $\Lambda$ probability density $\rho_{\Lambda}=(\rho^{c}_{n} \rho_{{\sf eq.},N -n}(\beta))$ with $\rho_{{\sf eq.},N -n}(\beta)$ the equilibrium probability density on the remainder of the system  $\Lambda$ comprised of $(N-n)$ particles. Since the above $\rho_{\Lambda}$ is stationary, it commutes the full system Hamiltonian $H$  on $\Lambda$
(similar to Eq. (\ref{comm})). Along similar lines, we may also construct a stationary $\rho_{\Lambda}$ that is the product of $(N/n)$ stationary probability densities of the form of $\rho^{c}_{n}$. If ensemble equivalence holds including the Eigenstate Thermalization Hypothesis \cite{ETH1,ETH2} then our earlier derivation of Section \ref{sect:proof_eqpp} can be repeated for any one of these forms for the above stationary $\rho_{\Lambda}$ to establish that $\rho^{c}_{n}$ can be written in the form of Eq. (\ref{pp}).

   \subsection{Corrections due to finite lifetime of the metastable state}
   \label{finite_time}

   In Section \ref{sect:proof_eqpp}, we examined systems for which taking divergent averaging time of the reduced probability densities ($\tau \to \infty$) in Equation (\ref{rlta}) did not yield any discernible difference as compared to the instantaneous probability densities. We now briefly shift our focus towards finite-time corrections to the metastable state and estimates of the ``near-stationarity'' of this state. Although the metastable state of the supercooled fluid is, to a very good approximation, stationary, it is not  devoid of temporal fluctuations. Eqs. (\ref{full},\ref{trivw},\ref{rlta},\ref{ft}) merely outlined fundamental general relationships. To assess the experimental relevance of Eq. (\ref{pp}) and establish a connection with the physics we wish to explore, we emphasize the empirical observation that local observables in the supercooled liquid appear stationary within the time range of $t_{\min} \le t \le t_{\sf xtal}$.

To capture this observed stationarity, we may set the time interval $\tau$ in Eq. (\ref{rlta}) to be 
$(t_{\sf xtal} - t_{\min})$. Specifically, in Eq. (\ref{rlta}), we consider times $ 
(t_{\sf xtal} - t_{\min}) \ge t' \equiv (t - t_{\min}) \ge 0$. 
Empirically, for these times $t'$, 
the local few body (i.e., $n$-body with small finite $n$) reduced probability density of the supercooled fluid is static and thus equal to its long-time average, given by
\begin{eqnarray}
\rho_{{\sf s.c.},n} = \rho_{n,l.t.a}.
\end{eqnarray}
Here, $\rho_{n,l.t.a.}$ is evaluated for the prescribed large (yet finite) averaging times $\tau$. Here, the maximal averaging times $\tau$
for which the time averaged reduced probability density does not change significantly can be viewed as a proxy for the lifetime ($\tau_{\ell}$) of the nearly stationary non-equilibrium system.

In the $\tau \to \infty$ limit, the time dependence of the reduced probability density is integrated out, rendering its long-time average identically stationary. However, for finite $\tau$, upon differentiating Eq. (\ref{rlta}), we obtain
\begin{eqnarray}
\label{derivative_lta}
\frac{\partial}{\partial \tau} \rho_{n,l.t.a}(\tau) = \frac{\rho_n(\tau) - \rho_{n,l.t.a.}(\tau)}{\tau}.
\end{eqnarray}
Given that the reduced probability density $\rho_n$ is bounded, it is evident that the right-hand side of Eq. (\ref{derivative_lta}) tends to zero, as it must, in the limit $\tau \to \infty$.

It follows from Eq. (\ref{derivative_lta}) that there are ${\cal{O}}(1/\tau)$ corrections to the right-hand side of Eq. (\ref{comm}). These are smaller by a dimensionless factor of $t_{\sf micro.}/\tau$ relative to the natural microscopic rate of change for single-atom dynamics. That is,
the long time average of Eq. (\ref{rlta}) of fluctuations occurring at a frequency $\omega_{\sf micro.} = 2 \pi/t_{\sf micro.}$ will lead to corrections that scale as ${\cal{O}}(t_{\sf micro.}/\tau)$. If $\tau$ is much longer than the natural timescale $t_{\sf micro.}$ for microscopic atomic dynamics time scales, such as the mean free time (as is typically the case for supercooled fluids), then these changes may be relatively insignificant when considering the form of the reduced few particle probability density in Eq. (\ref{pp}). For instance, in supercooled liquids, the ratio of the inter-atomic spacing to the speed of sound, which provides an estimate of the oscillation time (magnitude of $t_{\sf micro.})$, could be on the order of $10^{-13}$ seconds. The ultimate crystallization time $t_{\sf xtal}$ (the pertinent scale of $\tau$ in Eq. (\ref{derivative_lta})) is far larger. 
The precise value of the crystallization time depends on the material and temperature. Silicate and certain bulk metallic systems may take up to several minutes to crystallize (and be of the order of $\sim 10^{2}$ seconds and beyond just above the glass transition temperature $T_{g}$ or below the equilibrium melting temperature) \cite{Kelton_review,Ganorkar}. Glycerol solutions may remain supercooled for over $10^{4}$ seconds just above 190K \cite{glycerol-TTT}. Other supercooled liquids may avoid crystallization for few seconds or less
(the shortest times in some metallic glass formers being $\sim 10^{-3}$ seconds \cite{Kelton_review,Ganorkar}).
For different materials, the times $t_{\sf xtal}$ over which the supercooled liquid avoids crystallization are provided in the Time-Temperature-Transformation (TTT) curves. In all cases, $t_{\sf xtal}$ is many orders of magnitude larger than $\tau_{\sf micro.}$. A detailed account of the nucleation processes in supercooled fluids is given in excellent textbooks \cite{KG}.

Plugging in these empirical upper and lower bounds for $\tau_{\sf xtal}$ in supercooled fluids and setting in Eq. (\ref{rlta}) the averaging time to be  $\tau \sim t_{\sf xtal}$, one may anticipate that, across all temperatures of the supercooled liquid, the corrections to Eq. (\ref{pp}) scale as $10^{-10} \gtrsim (t_{\sf micro.}/t_{\sf xtal
}) \gtrsim 10^{-17}$. \\

We next very briefly review why, physically, $t_{\sf xtal}$ may be so large by comparison to natural microscopic time scales. As is well known, the system can persist in the supercooled state for an extended period due to either: \\

(I) A very low diffusion constant and high viscosity at low temperatures (resulting in a prevalence of solid-like states in the reduced equilibrium probability density for the supercooled liquid) \\

or \\

(II) A very small number of nucleation sites at high temperatures just below those of equilibrium melting (owing to the dominance of equilibrium fluid-type states in the reduced equilibrium probability density in this regime). \\

As a consequence of these two combined pincer-like effects, the crystallization rate remains extremely low (with the system being ``stuck'' in the metastable supercooled liquid state) at all temperatures below that of equilibrium melting. This leads to the said  value of the crystallization time  $t_{\sf xtal}$ to be many orders of magnitude larger than those associated with microscopic atomic motion, which are set by the temperature itself in semi-classical systems. The minimal crystallization time typically occurs between the two diametrically opposite  low and high temperature limits of (I) and (II). In Section \ref{sec:dynamics}), we will briefly discuss the bottleneck of  item (I) as it relates to the viscosity derived within our framework.

 This sluggish crystallization rate  suggests an ``adiabatic'' description of the nearly stationary reduced probability density $\rho_{n}$ in which the conditional probability distribution $P$ in Eq. (\ref{pp}) may accordingly acquire a weak time dependence. Indeed, regardless of the time ${\sf t}$ about which it is centered, whenever the reduced probability density exhibits no noticeable change over a long time window of width $\tau$, the proof of Section \ref{sect:proof_eqpp} will apply. 
This, in turn, mandates that, for nearly static systems, the average of the reduced probability density over such a time interval must be of a form similar to that of Eq. (\ref{pp}) yet with a conditional probability $P_{\sf t}(\beta, \mu, \ldots| \beta', \mu', \ldots)$ that may, in principle, change slowly with ${\sf t}$ \footnote{As they must, the large $\tau$ averages of the reduced probability densities $\rho_{n,l.t.a.}$ and hence the conditional probabilities $P_{\sf t}$ are nearly constant in ${\sf t}$ whenever all few body observables exhibit no appreciable dynamics as is the case for supercooled liquids in their long lived metastable state.}. In Section \ref{locality} and in the Appendix, we will discuss systems with local correlations and illustrate how locality mandates a temporal evolution. A weak time dependence of $\rho_{n}$ is consistent with the said ``adiabatic'' conditional probability distributions $P_{\sf t}$ which are not completely stationary and may fluctuate very slowly with time. We will turn to a qualitative discussion of the time dependence of effective local state variables in Section \ref{hyd}. Henceforth, we will largely omit the explicit ${\sf t}$ subscript of the conditional probability $P$.


\subsection{Spatially local near-stationary reduced probability densities}
\label{locality}

The sole assumptions that we employed in in our simple proof of the theorem of Eq. (\ref{pp}) were those of stationarity of the reduced probability density and ensemble equivalence (including the stronger assumption of the Eigenstate Thermalization Hypothesis in the case of a non-smooth ${\cal P}$). In particular, it is notable that in our above illustration of Eq. (\ref{pp}) as representing the most general stationary reduced probability densities, we {\it did not} invoke locality. As we now explain, for non-local probability densities, the requirement of being stationary is at odds with that of having local clustered spatial correlations (i.e., of the covariance decaying to zero for increasing interparticle distances). This is an important point since, conventionally, the reduced probability densities $\rho_{{\sf eq.}, n}$ of equilibrium systems governed by local Hamiltonians $H$ feature clustered correlations. We next turn to the consequences of the analog of Eq. (\ref{pp}) if it were valid for non-local reduced probability densities. To illustrate the basic premise, we consider the simplest nontrivial example- that of two particle reduced probability densities. To avoid an overly cumbersome notation, in what follows we will omit state variables other than $\beta$ and $\beta'$ and employ the shorthand $P(\beta|\beta')$ for the conditional probability in Eq. (\ref{pp}). For observables $A$ and $B$ that are, respectively, associated with single particles at spatial positions ${\bf{x}}$ and ${\bf{y}}$, the associated covariance (the connected correlation function) in general equilibrium system typically vanishes,
        \begin{eqnarray}
        \label{equilshort}
         && G_{{\sf eq.}, {AB}}({\bf{x}}, {\bf{y}})   \equiv \Big({\sf Tr}(\rho_{{\sf eq.}(\beta'), 2}AB) \nonumber
           \\ && - ({\sf Tr}(\rho_{{\sf eq.},1}(\beta')A))({\sf Tr}(\rho_{{\sf eq.},1}(\beta')B) \Big)  \to 0,
        \end{eqnarray} 
        as $|{\bf{x}}- {\bf{y}}| \to \infty$. From this, it immediately follows that for general
        $P(\beta| \beta')$, the associated states of Eq. (\ref{pp}) (which as we have proven above capture all possible stationary reduced probability densities regardless of the spatial separation between the particles) may exhibit long-range correlations. The demonstration of this assertion is straightforward. Rather explicitly, for a general fixed inverse temperature $\beta$ and arbitrary spatial separations  $|{\bf{x}}- {\bf{y}}|$, the covariance in a general stationary non-equilibrium system (for which Eq. (\ref{pp}) applies),
        \begin{eqnarray}
        \label{non_equil_long}
       &&  G_{{\bf neq},AB}({\bf x}, {\bf y}) \nonumber
       \\ && \equiv {\sf Tr}(\rho_{2}(\beta)AB) - ({\sf Tr}(\rho_1 (\beta) A))({\sf Tr}(\rho_{1}(\beta)B)) \nonumber 
        \\ && = \int d\beta' P(\beta| \beta') {\sf Tr}(\rho_{{\sf eq.},2}(\beta') AB) \nonumber
        \\ && - \int d\beta' P(\beta| \beta') {\sf Tr}(\rho_{{\sf eq.},1}(\beta') A) \nonumber
        \\ && \times \int d\beta' P(\beta| \beta') {\sf Tr}(\rho_{{\sf eq.},1}(\beta') B) \nonumber
        \\ && = \int d\beta' 
        \Big( P(\beta| \beta') {\sf Tr}(\rho_{{\sf eq.},1}(\beta') A ) {\sf Tr}(\rho_{{\sf eq.},1}(\beta') B ) \Big)\nonumber
        \\&&  - \int d\beta'  P(\beta| \beta') {\sf Tr}(\rho_{{\sf eq.},1}(\beta') A) \nonumber
        \\ && \times \int d\beta'  P(\beta| \beta') {\sf Tr}(\rho_{{\sf eq.},1}(\beta') B),
        \end{eqnarray}
        where, in the second line,  $\rho_1$ and $\rho_2$ respectively denote the reduced single and two-particle probability densities in the non-equilibrium system. Here, we inserted Eq. (\ref{equilshort}) in the second equality. As Eq. (\ref{non_equil_long}) indeed makes clear, unless $P(\beta|\beta')$ is a delta function in $\beta'$ (as, e.g.,  discussed in Eq. (\ref{different_is_more}) for general non-equilibrium systems)  then contrary to the clustered correlations of Eq. (\ref{equilshort}), the covariance between arbitrarily far separated observables $A$ and $B$ with $|{\bf{x}}- {\bf{y}}| \to \infty$        
        need not vanish identically,
        \begin{equation}
\label{eq:non-local}
        \lim_{|{\bf{x}} - {\bf{y}}| \to \infty} G_{{\sf neq}, AB}({\bf x}, {\bf y}) \neq 0.
        \end{equation}
        To ensure clustering in its minimal form for two-particle correlations, for asymptotically far separated sites ${\bf{x}}$ and ${\bf{y}}$, the reduced probability density $\rho_{2}({\bf{x}}, {\bf{y}})$ must, instead of Eq. (\ref{pp}), become equal to the product of single particle probability densities, $\rho_{1}({\bf{x}}) \rho_{1}({\bf{y}})$. Given our earlier demonstration that Eq. (\ref{pp}) is the most general reduced probability density, such a deviation implies that the reduced few particle probability density 
        in a system in which clustering of correlations is obeyed will {\it not remain time independent}. That is, taken together, Eqs. (\ref{pp}, \ref{eq:non-local}) imply that unless the system is in a state of equilibrium associated with its Hamiltonian (or is, effectively, in an state  equilibrium of another Hamiltonian such as the ferromagnet discussed in Section \ref{sect:proof_eqpp}), a system having spatially clustered correlations may be near-stationary with weak time dependence yet its reduced probability density cannot be completely time independent. 
That is, for such non-equilibrium systems,  
{\it clustered correlations imply non-stationarity.} 
In view of this maxim, one might anticipate that when the spatial scale on which violations of the general stationarity solution of Eq. (\ref{pp}) are apparent becomes shorter (i.e., the more spatially clustered the correlations are), the temporal fluctuations of the associated reduced few particle probability densities $\rho_{n}$ may  become larger. In Appendix \ref{example=cluster}, we 
discuss an approximate simple pedagogical example for which this maxim may vividly come  to life. Far more broadly, if an inverse correlation length $\xi^{-1}$ 
associated with a two point covariance is a function of the inverse lifetime $\tau_{\ell}^{-1}$ of the non-equilibrium state and if $\xi^{-1}$ has a well defined $\tau_{\ell} \to \infty$ limit (where Eq. (\ref{eq:non-local})  needs to generally apply following Eq. (\ref{non_equil_long})) then, in this asymptotic limit, $\xi^{-1}$ must trivially   \footnote{This explicitly  follows from (i) the positivity of the correlation length $\xi$ and lifetime $\tau_{\ell}$, (ii) the divergence of the correlation length  $\xi$ implied by Eq. (\ref{eq:non-local}) when the lifetime of the non-equilibrium system $\tau_{\ell} \to \infty$, and the above noted (iii) assumed well-defined function $\xi^{-1}(\tau^{-1}_{\ell})$ for asymptotically large $\tau_{\ell}$. Property (ii) implies that $\lim_{\tau_{\ell} \to \infty} \xi^{-1}(\tau_{\ell}^{-1}) =0$. Property (i) then  mandates that in the vicinity of the latter asymptotic   $\xi^{-1}=\tau_{\ell}^{-1}=0$ point, the function  $\xi^{-1}(\tau_{\ell}^{-1})$ must be monotonically non-decreasing.} be strictly monotonically non-decreasing in the inverse lifetime  $\tau_\ell^{-1}$. 
Thus, Eq. (\ref{pp}) and its extensions may describe {\it completely stationary} states with a $P$ that is not of a delta function form only for {\it local} few  body states. In Appendix \ref{nearly-commute}, we briefly consider quantum systems. 
 As we further discuss in Appendix \ref{ap:approx}, taking into account  clustering, one may consider  approximations to the probability density that are static. Whenever the system exhibits clustering, the energy density and all other intensive quantities become sharp.

\subsection{Hydrodynamic and mechanical perspectives on local equilibration}
\label{sub:hydro}

When the liquid is rapidly supercooled, it does not ``have enough time'' to veer towards its maximal entropy global equilibrium configuration. Instead, it may only equilibrate to achieve the said local stationary state. In this state, short and medium range atomic orders appear and have been characterized. These resulting structures are spatially nonuniform. Importantly, such nonuniform spatial structures leads to large (by comparison to those found in equilibrium) fluctuating local potential energies and force fields and consequent large fluctuations for dynamic and other properties. Microscopically, a distribution of local temperatures as in Eq. (\ref{pp}), and thus of the effective temperature as it appears in the globally averaged few body probability densities, is inevitable in, e.g., in Euler hydrodynamics (as well as diffusive processes) that we briefly review next. The aim of this subsection is to make the considerations of a distribution of state variables that is relevant to local observables more intuitive. 

\subsubsection{Hydrodynamic motivation}
\label{hyd}
It is not a priori clear that supercooled liquids may be described by conventional hydrodynamics. Memory function approximations and mode coupling effects have been heavily investigated with higher order closure schemes leading to progressively more accurate results \cite{mode_coupling}. Mode coupling has seen notable successes yet, in its common form, misses essential features of the phenomenology of supercooled liquids \cite{mode_coupling}. Nonetheless, in order to obtain a qualitative simple and vivid analog of our approach, we briefly regress to discussing some basic aspects of simple hydrodynamics.  

If a charge density $q$ is associated with a global conserved quantity (charge) 
then $\partial_{t} q + \nabla \cdot {\bf j} =0$ with ${\bf j}$ the respective current  density. For illustrative purposes, we  consider the set of all conserved charges, with densities labelled by $q_a$, in a one-dimensional system, with the flux Jacobian 
\begin{eqnarray}
A_{a}^{b} \equiv \frac{\partial \langle j_a \rangle}{\partial \langle q_b \rangle}.
\end{eqnarray}
The respective Euler equations then read
\begin{eqnarray}
\label{Euler}
\partial_{t} \langle q_a \rangle + A_{a}^{b} \partial_x \langle q_{b} \rangle =0.
\end{eqnarray}
The local expectation values $\langle q_{a}\rangle$ are those obtained in the canonical ensemble with the substitution of the state variables by their local values $\beta'(x,t), \mu'(x,t), \cdots$. The number of the latter (local equilibrium) state variables $\beta', \mu', \cdots$ is generally equal to the number of independent globally conserved quantities  (volume integrals of $\{q_{a}\}$). We will therefore have as many Euler type equations as local state variables.  Since, in equilibrium, an expectation value $\langle q_{a} \rangle$ is a function of the state variables $\beta', \mu', \cdots,$ Eqs. (\ref{Euler}) lead to equations associated with the spatial gradients of the local equilibrium temperature, local chemical potential, and any other state variables. In general, the equilibrium $\beta', \mu', \cdots$ are functions of space (and time). Spatially fluctuating local temperatures and chemical potentials appear in numerous fields- e.g., in descriptions of bubble nucleation \cite{superheat} and plasma physics
\cite{localeTplasma}.

Broader than Eqs. (\ref{Euler}), an assumption of hydrodynamics in general dimensions \cite{Doyon} is that
the expectation values of all quasilocal observables $w$ are those evaluated with a Gibbs distribution defined by local equilibrium state variables
\begin{eqnarray}
\label{local-hydro}
 \langle w({\bf x},t) \rangle \approx \langle w({\bf 0},0) \rangle_{\beta'({\bf x},t), \mu'({\bf x},t) , \cdots}.   
\end{eqnarray} 

Thus, the reduced probability density $\rho_{\sf eq.}({\bf x},t)$ for any local region centered about $({\bf x},t)$ is given by the equilibrium Boltzmann distribution with its respective state variables. Although these local coarse grained  state variables $\beta'({\bf x},t), \mu'({\bf x},t) , \cdots$ may, in particular, vary with the time $t$, the entire distribution of these values (i.e., their frequency over the entire volume of the system as, e.g., seen when binned in a histogram of these values constructed at any fixed time $t$) can be stationary. Indeed, the latter distribution is time translationally invariant in the effectively equilibrated state of  any liquid in which 
Eq. (\ref{local-hydro}) empirically holds for all local observables $w$. Such a distribution may be identified with the conditional probability $P$ of Eq. (\ref{pp}).  In the absence of long range correlation within a coarse grained hydrodynamic description of the system, one may describe the single and other few body local reduced probability densities via Eq. (\ref{pp}) with $P$ representing, as in the above, the relative frequency of the local state variables. As these state locally fluctuating state variables $\beta'({\bf x},t), \mu'({\bf x},t) , \cdots$ evolve according to the Euler equations (or other equations that further include diffusive and additional effects), their relative binned frequency over the entire system may also change in time (even if this frequency varies exceedingly slowly in time as in the case of the crystallization of a supercooled liquid discussed in Section \ref{finite_time}). In such situations, few body local expectation values will still be given by Eq. (\ref{pp}) yet, similar to our earlier discussion in Section \ref{finite_time}, the conditional probability distribution $P$ in the integrand of Eq. (\ref{pp}) may be of  an ``adiabatic'' form $P_{\sf t}$ in which the dependence  on the equilibrium state variables $\beta', \mu', \cdots$ varies slowly with the time ${\sf t}$.

\subsubsection{A one-dimensional mechanical system}
\label{1d-decoupled_particles}

We now highlight an exceptionally simple system where (even when this system is arbitrarily far away from equilibrium) the distribution $P$ of local state variables will remain exactly stationary at all times. Similar to Section \ref{decoupled_solvable}, this system exhibits a trivial evolution. This example exhibits a stationary probability density of the local temperatures $\beta'$ when sampled over the entire system. As in Section \ref{hyd}, $\beta'({\bf x},t)$ may vary in space and time. When averaging over the entire system, the probability of obtaining a particular $\beta'$ does not change as the system evolves in time. Towards that end, we consider equal mass particles colliding elastically along a one-dimensional chain. In these one dimensional elastic collisions, any pair of colliding particles simply exchange their velocities ($v_{i}, v_{j} \to v_{j}, v_{i}$). Thus, in the absence of coupling to an external environment, the global velocity distribution of the particles (and thus the distribution $P(\beta|\beta')$ of the effective ``local inverse temperatures'' $\beta'$ associated with these kinetic energies) in such a one dimensional system remains trivially (and permanently) stationary as a function of time. This holds true for {\it any} distribution of initial local velocities or $\beta'$ values (including those arbitrarily far away from a Maxwell-Boltzmann distribution of the velocities or, equivalently, a delta function function distribution in $\beta'$).
Similar to the example of Section \ref{decoupled_solvable}, the above one-dimensional system of colliding elastic particles is, clearly, somewhat special. More generally, as discussed in Section \ref{locality}, finite time metastable states may appear.

\subsection{Consequences of our central relation- predictions for general local observables}
\label{consequences}

Eqs. (\ref{pp}, \ref{pp'}) trivially lead to a very strong prediction: 
When averaged over the entire system $\Lambda$, {\it all few body observables} $Q$  
(including, e.g., those whose averages yield the correlation functions and local structural measures) in the stationary 
are trivially related to those in the truly equilibrated  
system at different temperatures, chemical potentials, ... (see the first integral below) and energy densities, number density, ... (second integral), 
\begin{eqnarray}
\label{Qavglong}
 \! \! \! \! \! \! \! \langle Q \rangle_{\sf stationary}  =  && \int d\beta' d \mu'  \ldots \Big( P(\beta,\mu, \ldots| \beta',\mu
',\ldots) \nonumber 
\\
&& \times \langle Q \rangle_{{\sf eq.~}\beta',\mu',\cdots} \Big) \nonumber
\\ && \! \! \! \! \! \! \!+ \int_{\cal{PTEI}} d\epsilon'd {\tilde{\sl n}}' \cdots 
\Big( P(\epsilon, {\tilde{\sl n}}, \ldots| \epsilon',{\tilde{\sl{n}}}
',\ldots) \nonumber
\\ && \times \langle Q \rangle_{{\sf eq.~}\epsilon',{\tilde{\sl n}}',\cdots}\Big).
\end{eqnarray}
Here, $\langle Q \rangle_{\sf eq}$ is the equilibrium expectation value. The  $\cal{PTEI}$ regime (Eq. (\ref{pp'})) integral in Eq. (\ref{Qavglong}) is over those energy and number densities for which equilibrium phase coexistence may occur. 
The stationary system has state variables such as the inverse temperature $\beta$ or energy density $\epsilon$, while the equilibrium averages are, respectively, defined by $\beta'$ or $\epsilon'$. Within the $\cal{PTEI}$, the equilibrium expectation values may, e.g., not be uniquely determined by temperature. That is, as underscored when deriving Eq. (\ref{pp'}), when discontinuous phase transitions occur, there may be a broad range of energy densities $\epsilon'$ associated with a single fixed inverse temperature $\beta'$ of the equilibrium system. To account for this possibility, the $\cal{PTEI}$ contributions appear in Eqs. (\ref{pp'}, \ref{Qavglong}). The relation of Eq. (\ref{Qavglong}) was motivated in Refs. \cite{1parameter,longrange,PS} when considering a non-equilibrium static probability density.  As explained in  Section \ref{locality}, the considerations in the current work apply to nearly static local few body observables (when these are averaged over the entire system), not to ones associated with non local correlation functions. Formally, the average in Eq. (\ref{Qavglong}) emulates that in spin glass problems \cite{SG1} with the quenched disorder average over coupling constants replaced by the conditional probability $P$ for the state variables. \bigskip

It is important to underscore that in Eq. (\ref{Qavglong}), the {\it very same conditional probability} $P$ universally appears for the expectation values of all few body observables $Q$. This is a strong prediction of the theory.
\bigskip

In the context of the possible experimental ramifications that we focus on in the current work, a nontrivial probability distribution $P$ must connect averages $\langle Q \rangle_{\sf s.c.}$ the supercooled liquid with averages $\langle Q \rangle_{\sf eq.}$ in its truly equilibrated counterpart. We emphasize that $P$ cannot be a delta function for otherwise the system could not be differentiated from an equilibrated solid or fluid at the same temperature. A distribution of $\beta'$ values implies that {\em{the average single and few body dynamics}} may display {\em heterogeneous} behaviors (similar to those observed in experiments and simulations, e.g., \cite{Cavagna,BB,DH1,DH2,DH3,rotation}) by the above necessary need of superposing different temperatures of the equilibrium system. This does not mean that heterogeneous dynamics may not appear for non local observables- such dynamics are simply not a consequence of our framework. 

An important corollary of Eq. (\ref{Qavglong}) is that singularities in various observables that appear in equilibrium phase transitions can now be ``smeared out'' by the conditional probabilities $P$. Nonetheless, the values of the  temperature, chemical potential, and of  other state variables at which equilibrium phase transitions occur (and at which $\langle Q \rangle_{{\sf eq.~}\beta',\mu',\cdots}$
exhibit singularities) may still play a prominent role. We will discuss how this might come to life in Section \ref{sec:exp'} where we will discuss the viable appearance of the equilibrium melting (or liquidus) temperature in describing the temperature dependence of the dynamics and observables in supercooled liquids.

\subsection{Effective many body probability densities}
\label{sec:eff}

 A given reduced probability density $\rho_{n, l.t.a.}$ may be expressed as the partial trace of numerous many body probability densities $\rho_{\sf eff}$ other than the long-time average  
         $\rho_{\Lambda, l.t.a.}$ of Eq. (\ref{llta}). That is, we can write $\rho_{n,l.t.a}$ as a partial trace of such effective probability densities,
         \begin{eqnarray}
         \label{eff_defn}
            \rho_{n, l.t.a} = {\sf Tr}_{n+1, \cdots, N} ~ \rho_{\sf eff}. 
         \end{eqnarray}
         In general, there are indeed multiple viable $\rho_{\sf eff} \neq \rho_{\Lambda, l.t.a}$ that will yield the same reduced probability density $\rho_{n, l.t.a}$.
         Not all of these effective reduced probability densities will adhere to clustering conditions. However, they will all yield the same reduced probability density. Explicit examples of effective $\rho_{\sf eff}$ can be constructed for the models of Section \ref{decoupled_solvable}. In general, a simple recipe (already implicit in our derivation of Eq. (\ref{pp}) in Section \ref{sect:proof_eqpp})) may be provided for some such effective probability densities. Similar to Sections \ref{decoupled_solvable} and \ref{locality}, in what follows, we will, for simplicity of notation consider only a single state variable- that of the inverse temperature. The reduced single particle probability density in the decoupled (``${\sf dec.}$'') model example of Eq. (\ref{trivexample}) can, e.g., be written as 
\begin{eqnarray}
 \rho_{1} =    {\sf Tr}_{2,3, \ldots N}~(\rho^{\sf dec.}_{{\sf eff}, \Lambda}).
 \end{eqnarray}
 Here, the effective many body probability density $\rho^{\sf dec.}_{{\sf eff}, \Lambda}$ is given by a weighted sum (or integral) of the thermal equilibrium many body system,
 \begin{eqnarray}
 \label{eff_example_1}
  \int d \beta' ~P(\beta|\beta')~ \frac{e^{-\beta' H}}{Z(\beta')},
\end{eqnarray}
with $H$ the system Hamiltonian and the equilibrium partition function
\begin{eqnarray}
\label{eff_example_2}
Z(\beta') = {\sf Tr}(e^{-\beta' H}).
\end{eqnarray}
The effective many body probability density of Eqs.  (\ref{eff_example_1},\ref{eff_example_2}) goes beyond reproducing the reduced  probability for the example of Eq. (\ref{trivexample}). Indeed, Eqs.  (\ref{eff_example_1},\ref{eff_example_2})  more generally yield all order (i.e., arbitrary $n$) reduced probability densities of the form of Eq. (\ref{pp}) of a given $P$. For a distribution $P(\beta|\beta')$ that has a non-vanishing standard deviation of its $\beta'$ values, the standard deviation of the system energy (i.e., that of the Hamiltonian $H$) as computed with Eq. (\ref{eff_example_1}) is extensive. 

We emphasize that a given (static or near static) reduced probability density that is thus equal to its long time average and (either exactly or approximately) given by Eq. (\ref{pp}) may correspond (Eq. (\ref{eff_defn})) to multiple many body effective probability densities. The latter effective probability may have finite energy density fluctuations (as in, e.g., Eq. (\ref{eff_example_1})) or, as we next underscore, no fluctuations whatsoever of the energy density. 
To illustrate this point, we note that in addition to the effective probability density of $\rho^{\sf dec.}_{{\sf eff}, \Lambda}$, other effective probability densities can be trivially constructed in which the full Hamiltonian $H$ in
Eqs. (\ref{eff_example_1}, \ref{eff_example_2}) is replaced by
a product of $N$ single particle ($n=1$) probability densities associated with $H_1$ (i.e., $\rho_{\Lambda}(\beta)$
of Eq. (\ref{trivexample})) or with the Hamiltonian $H$ in Eqs. (\ref{eff_example_1}, \ref{eff_example_2}) replaced by a sum of two local Hamiltonians $(H_1+H_2)$, or by the sum of three such local Hamiltonias, etc. For each of these sums of local Hamiltonians, the probability densities are different yet the associated reduced single ($n=1$) particle probability density is the same. In general, given any reduced probability $\rho_n$ one may find a minimal size $n-$ body Hamiltonian $H_n$ such that $\rho_n$ can be equated to the Boltzmann probability density associated with $H_n$. For finite $n$, the fluctuations of all intensive quantities will vanish in the thermodynamic limit. All of these effective probability densities will yield identical expectation values for all local few body observables.

         Although there are numerous $N-$body effective probability densities $\rho_{\sf eff}$ satisfying Eq. (\ref{eff_defn}), in the general case (as we previously illustrated when examining the viable form of $\rho_{\Lambda, l.t.a.}$ in the $\tau \to \infty$ limit) only those probability densities that are of the form of the righthand side of Eq. (\ref{longpp+}) (which are the same as those given by 
the effective probability density of Eq. (\ref{eff_example_1})) 
are completely stationary under evolution with $H$. As we emphasized earlier, the clustered states $\rho_{\Lambda, {\sf close.}}$ of Eqs. (\ref{clustering_rel},\ref{pp_prod}) are only approximations to the full many body probability density. As is evident from Eq. (\ref{longpp+}), the sole many body probability densities that are completely stationary are given by weighted average of equilibrium probability densities appearing in the effective probability densities of Eq. (\ref{eff_example_1}). Furthermore, as we emphasized earlier, states of the form of Eq. (\ref{r2w})  will (unless $\xi \to \infty$) exhibit dynamic fluctuations at sufficiently long times. Thus, in general, only reduced probability densities of the form of Eq.(\ref{different_is_more}) can be associated with completely stationary many body probability densities that obey clustering.       
Since both $\rho_{\sf eff}$ and $\rho_{\sf eq}$ are normalized probability
distributions, we can readily relate to the two via a kernel- the conditional probability $P(\beta|\beta')$. 
By virtue of being a function (henceforth denoted by $f$) of the Hamiltonian $H$, the distribution $\rho_{\sf eff}$ is related to $\rho_{\sf eq.}$ via a simple Laplace transform. 

Extending the general notions underlying Eqs. (\ref{eff_example_1}, \ref{eff_example_2}), let us now return to $\rho_{\sf eff.}$ and express it in terms of a Laplace transform of the latter equilibrium distribution, 
\begin{subequations}
\begin{align}
\rho_{\sf eff} = f(H) = {\cal{L}}(\tilde{f}) =  \int d \beta' ~\tilde{f}(\beta') ~e^{-\beta' H}, \label{fL} 
\\  
\tilde{f}(\beta') = {\cal{L}}^{-1}(f) =  \frac{1}{2 \pi i} \lim_{w \to \infty} \int_{\gamma - i w}^{\gamma + t w} dz~ f(z)~ e^{z \beta'}. \label{Lf}
\end{align}
\end{subequations}
 Here, ${{\cal L}}$ and ${\cal{L}}^{-1}$ denote the Laplace transform and its inverse. Thus, up to a rescaling (set by the partition function $Z(\beta')$ of Eq. (\ref{eff_example_2})), the inverse
Laplace transform $\tilde{f}$ may be regarded as the conditional probability $P(\beta|\beta')$
that the effective equilibrated system (described by $\rho_{\sf eff}$) is at an inverse temperature
 $\beta$ given that the equilibrium system is known to be at an inverse temperature $\beta'$. 
 
 If, for any fixed $\beta$, the conditional probability $P(\beta|\beta') = \tilde{f}$ is not a delta function in $\beta'$ then it may  generally describe a system with a finite variance of the energy density and thus display long range correlations.
 That is, in such instances, the correlation function $G_{ij}$ of Eq. (\ref{sE1}) must remain finite for arbitrarily large separations $|i-j|$ since the lefthand side of Eq. (\ref{sE}) will, generally, not vanish. Nonetheless, given Eq. (\ref{eff_defn}), long time averages of few body local expectation values computed with the effective probability density of Eq. (\ref{fL}) will be equal to those in the system with only local (clustered) correlations. Although we will largely abstain from doing so, the expectation values that we will next compute in Section \ref{sec:exp'} may be formally obtained by computing expectation values with such effective many body probability densities.

\section{Applications to Experimental data: predictions and estimates}
\label{sec:exp'}

In Section \ref{sec:stationary+near-stationary}, we set some groundwork in deriving Eqs. (\ref{pp}, \ref{pp'}), explaining the physics behind these relations, and outlining some of the predictions can be made with these relations. We briefly suggested (Section \ref{finite_time}) how the finite time lifetime of the supercooled fluid state might lead to exceptionally small corrections.  We now describe how various experimental data may be analyzed with the aid of these equations. Our focus continues to be on supercooled fluids for which we will further motivate a particular (scale free Gaussian) form of the conditional probability $P$ \footnote{In this application (and others like it), the lower $t'=0$ limit of the integrals  corresponds, as in Section \ref{finite_time}, to the minimal waiting time $t=t_{\min}$ for the system to reach its long-lived nearly stationary state.}. As we will explain, however, if some experimental data are available (e.g., measurements of the single particle velocity probability density) then predictions concerning the viscosity can be made with no additional assumptions.

\subsection{Extracting the effective conditional probabilities $P(\beta|\beta')$ from measured velocity distributions}
\label{sec:extract}

 As we now demonstrate, a prediction of our approach is that of a non-trivial distribution of velocities. Towards this end, we recall that regardless of the specifics of the potential energy and possible phase coexistence, given an inverse temperature $\beta'$, in all equilibrium classical non-relativistic systems adhering with separable (potential energy) position and momenta (kinetic energy) dependence, the probability density associated with any Cartesian velocity component $v_{i \alpha}$ of a given particle $i$ of mass $m_i$ is $\sqrt{
 \frac{2 \pi }{m_{i} \beta'}} ~e^{-\beta' m_i v^2_{i \alpha}/2}$ (i.e., the standard Maxwell-Boltzmann distribution). Eqs. (\ref{pp}, \ref{pp'}) then assert that in the supercooled system,
\begin{eqnarray}
\label{no_Maxwell_dist}
P_{\beta}(v_{i \alpha}) =  \int d \beta^{'} ~\sqrt{
 \frac{2 \pi }{m_{i} \beta'}}  ~e^{-\beta' m_i v^2_{i \alpha}/2} ~~P(\beta|\beta'). 
\end{eqnarray}
That is, at any fixed inverse temperature  $\beta$, the velocity distribution $P_{\beta}(v_{i \alpha})$ is a ``Gaussian transform'' \cite{Alecu} of the conditional probability $P(\beta|\beta')$.
Eq. (\ref{no_Maxwell_dist}) may be tested experimentally. Regretablly, thermostats used in typical numerical simulations (e.g., the celebrated Nose-Hoover \cite{Nose,Hoover}, Andersen \cite{Andersen_thermostat}, and Langevin \cite{Langevin1,Langevin2} thermostats) are all guaranteed, by construction, to yield a Maxwell-Boltzmann distribution of the velocities. Thus, due the nature of these thermostats, Eq. (\ref{no_Maxwell_dist}) may not be seen numerically when such thermostats are used. An upcoming work will detail numerical results obtained when such thermostats are not employed \cite{Giorgi}. In a different arena, experimental observations of non-Maxwellian velocity distributions were indeed seen  \cite{non-Maxwell}. Eq. (\ref{no_Maxwell_dist}) implying a non-Maxwellian distribution of velocities for a non-delta function $P$ differs with the configuration energy landscape approaches \cite{Cavagna,BB,Goldstein} for which a Maxwellian distribution is typically assumed. 
In the latter energy landscape approach, the effective potential may differ from that present in the equilibrium system yet the kinetic term remains invariant if the temperature or energy density is uniform for all basins. 
By contrast, the picture that emerges in our description is more symmetric in the phase space coordinates of the system. That is, non-equilibrium fluctuations appear in both position dependent (potential energy) as well momentum dependent (i.e., kinetic) contributions to the total energy. Our below discussion of the translational velocity may be replicated to angular velocities (which, similar to the translational velocities, have also been experimentally observed to be heterogeneous \cite{rotation}). 

A trivial change of variables  renders Eq. (\ref{no_Maxwell_dist}) into a Laplace transform. Consequently, Eq. (\ref{no_Maxwell_dist}) may be readily inverted to deduce the conditional probability $P(\beta|\beta')$ from the experimentally measured velocity distribution $P_{T}(v_{i \alpha})$. In terms of the equilibrium temperatures $T'$, at any fixed temperature $T$, the conditional probability 
\begin{eqnarray}
\label{Pfromv}
P(T|T') = \sqrt{\frac{\pi}{2}} \big(\frac{m_i}{k_{B}}\big)^{5/2} (T')^{-3/2} \nonumber
\\ \times \Big[ {\cal L}^{-1}(P_{T}(v_{i \alpha})) \Big]_{s= \frac{m_i}{2 k_{B} T'}},
\end{eqnarray}
where, as in Eq. (\ref{Lf}), ${\cal L}^{-1}$ is the inverse Laplace transform.
 For our application, especially if the ${\cal PTEI}$ contributions
are insignificant, we may make experimental predictions for all local observables in the supercooled liquid $\langle Q \rangle_{\sf s.c.}$ (Eq. (\ref{Qavglong})). That is, from experimental measurements of the velocity distribution $P_{T}(v_{i \alpha})$, one may determine $P(T|T')$ via the inverse Laplace transform of Eq. (\ref{no_Maxwell_dist}). This may then subsequently provide predictions for all  $\langle Q \rangle_{\sf s.c.}$ (Eq. (\ref{Qavglong})) and general correlation functions of local observables.

The non-vanishing moments of the velocity in the supercooled system at an inverse temperature $\beta$ are
\begin{eqnarray}
\label{v-equil-moment}
\langle v_{i \alpha}^{2 {\sf k}} \rangle_{\sf s.c.} = \frac{(2 {\sf k}-1)!!}{m_i^{\sf k}} \int d \beta' ~ (\beta^{'})^{-{\sf k}} ~P(\beta|\beta').
 \end{eqnarray}

In Eq. (\ref{v-equil-moment}), for a general (i.e., non-delta function) $P(\beta'|\beta)$, the higher temperature (lower $\beta'$) contributions in the above integral are boosted by a relative factor of $(T'
)^{k}$. The supercooled system may thus, on average, appear to move and flow more than its equilibrium crystalline counterpart at the same temperature. The ratios of the velocity moments of the supercooled liquid
to those of the equilibrium system (i.e., to those evaluated for the Maxwell-Boltzmann distribution) when both are kept at a temperature $T$,
\begin{eqnarray}
\label{v-ratio}
\frac{\langle v_{i \alpha}^{2 {\sf k}} \rangle_{\sf s.c.}}
{\langle v_{i \alpha}^{2 {\sf k}} \rangle_{\sf eq.}}
= \frac{\int dT' ~ (T^{'})^{\sf k} ~P(T|T')}{{T^{\sf k}}},
 \end{eqnarray}
provide information on the distribution $P(T|T')$. For a distribution $P(T|T')$ that, for fixed $T$, is Gaussian (or is any other general even function) in the temperature difference $(T-T')$, the latter ratio is, trivially, one for ${\sf k}=1$ 
i.e., the kinetic energies of the supercooled and equilibrium systems are the same) while higher velocity moments may generally differ (being larger in the supercooled system). Our above predictions for the velocity distributions are stronger than current metrics for dynamical heterogeneities which typically involve four-point correlation functions, self part of the van Hove functions, and other conventional measures that touch on varying relaxation rates. The modified velocity distribution that we discuss here may further bolster dynamical heterogeneities \cite{Cavagna,BB,DH1,DH2,DH3,rotation} already observed with conventional measures when thermostats are used  (for which, as noted above, the velocity distribution is forced to be Maxwellian). 

In subsequent sections, we will consider
the scale-free normal distribution ansatz given by
\begin{eqnarray}
\label{cluster}
P(T|T')= \frac{1}{T \overline{A} \sqrt{2 \pi}} \exp \Big[-\frac{1}{2}\Big(\frac{T - T'}{T \overline{A}}\Big)^2 \Big], 
\end{eqnarray}
with a dimensionless constant $\overline{A}$. 
 Invoking Eq. (\ref{v-ratio}), it is trivially seen that, for this distribution, the non-vanishing moments of the velocity at a fixed temperature $T$ are given by
\begin{eqnarray}
\label{moment-v}
\langle v_{i \alpha}^{2 {\sf k}} \rangle_{\sf s.c.}  
= \langle v_{i \alpha}^{2 {\sf k}} \rangle_{\sf eq.} \sum_{\ell=0}^{\lfloor {\sf k}/2 \rfloor} {{\sf k} \choose 2 \ell} \overline{A}^{2 \ell} (2 \ell -1)!!.
\end{eqnarray}
At each temperature $T$, the velocity distribution $P_{T}(v_{i \alpha})$ may be  measured in experiment and/or simulation. If it is of the form of Eq. (\ref{cluster}), then for each such temperature $T$, the distribution 
$P(T|T')$ can be inferred (Eq. (\ref{Pfromv})). In a similar way, one might  apply related considerations to molecular rotation velocities, e.g., \cite{rotation}. More broadly, we may have a general distribution that can be expressed (in a non unique way) as a sum of such Gaussians centered about temperatures $c_n T$ with varying widths,
\begin{eqnarray}
\label{superposition}
P(T|T') = 
\sum_\ell \Big(\frac{w_{\ell}}{c_\ell T \overline{A}_\ell(T) \sqrt{2 \pi}} \nonumber
\\ \times \exp \Big[-\frac{1}{2}\Big(\frac{c_\ell T - T'}{c_\ell T \overline{A}_\ell(T)}\Big)^2 \Big] \Big),
\end{eqnarray}
with normalized weights $\sum_{\ell} w_{\ell} =1$. As we will discuss in Section \ref{hydro} (and in Eq. (\ref{localcv}) in particular), Eq. (\ref{superposition}) 
emulates a mixture of local clusters of varying sizes with a local temperature distribution peaked about 
$c_\ell T$. The weights $w_{\ell}$ may vary with the temperature. For a continuous distribution of Gaussians, the discrete sum in Eq. (\ref{superposition}) will be replaced by an integral.

\subsection{Local entropy maximization for the supercooled system}
\label{hydro}

In what follows, we further build on some of the notions discussed in Section \ref{sub:hydro}. Towards that end, we briefly review certain basic concepts underlying entropy maximization and then apply these to our case in which such maximization is only locally realized. Expanding, to harmonic order, $\log \rho_{\sf eff.}$ about the state of maximal entropy that is of fixed energy leads to textbook type Gaussian distribution for $\rho_{\sf eff.}$ similar to that of a finite size system that is in equilibrium.
We will ask what occurs when the system is effectively partitioned into decoupled ${\sf n}-$particle clusters. 
These clusters are proxies for local or medium range structures (including any ``defects'' and ``locally preferred configurations'' \cite{Ryan2}) in the supercooled liquid. We will think of ``${\sf n}$'' as the number of particles in these decoupled clusters in order to arrive at simple estimates for the possible energy distribution in the supercooled liquid. An approximate coarse grained representation of such clusters relates to the hydrodynamic description of Section \ref{hyd}. We reiterate that here, however, we will presume that these clusters are decoupled and, unlike the hydrodynamic description, not require smoothness of local fields across their boundaries. We must stress that, clearly, the supercooled liquid cannot, at all, be partitioned into decoupled clusters; this will be assumed here only in order to arrive at simple qualitative estimates. Within such an idealized picture, ${\sf n}$ may serve as a proxy for the largest number of particles $n$ in the supercooled fluid for which Eq. (\ref{pp}) applies for local observables. Our aim is to arrive at a possible distribution $P(T|T')$ for effective equilibrium temperatures $T'$ given that the supercooled system is at temperature $T$. As is well known, entropy maximization (occurring for an equilibrated ${\sf n}$-particle cluster) of the probability density of the energy leads to a normal distribution, 
\begin{eqnarray}
\label{Gauss}
P(\Delta \epsilon') = \frac{1}{\sqrt{2 \pi \sigma^2}}e^{-\Delta \epsilon'^2/(2 \sigma_{\epsilon}^2)}.
\end{eqnarray}
Here, $\epsilon'$ denotes the energy of the ${\sf n}-$particle system with $\Delta \epsilon'$ the fluctuation of its energy about its equilibrium average. The variance of such idealized finite size clusters 
$\sigma_{\epsilon}^2 = k_{B} T^2 c_v,$
where $c_v$ is the constant volume heat capacity of these local uncorrelated ${\sf n}-$particle regions. 

In Eq. (\ref{Gauss}),
\begin{eqnarray}
\Delta \epsilon' = \int_{T}^{T'} c_{v}(T'')~ dT'' 
\end{eqnarray}
is the local energy change of the local region at a local fluctuating temperature $T'$ with the global average temperature being $T$. We write 
\begin{eqnarray}
\label{av:c}
   \Delta \epsilon' = \overline{c} (T' - T)  
\end{eqnarray}
where $~\overline{c}~$ is the average constant volume local heat capacity of the cluster in the interval $[T',T]$ so that 
$\exp[-\Delta \epsilon'^2/(2 \sigma_{\epsilon}^2)] = \exp[-(\overline{c}(T-T'))^2/(2 k_{B}~ c_v(T)~ T^2) ].$   
Equivalently, 
\begin{eqnarray}
  \exp[-\Delta \epsilon'^2/(2 \sigma_{\epsilon}^2)] \sim 
  \exp \Big[-\frac{1}{2}\Big(\frac{T - T'}{T \overline{A}}\Big)^2 \Big],
  \end{eqnarray}
with the earlier discussed dimensionless parameter
\begin{eqnarray}
\label{localcv}
\overline{A}^2 = k_{B} \frac{c_{v}(T)}{\overline{c}^2}.
\end{eqnarray}
The effective distribution then becomes of the form of Eq. (\ref{cluster}).

To obtain order of magnitude estimates, consider $c_v \sim \overline{c} \sim (3{\sf n} k_{B})$ where ${\sf n}$ is the number of atoms in the locally fluctuating region. This implies that for, e.g., ${\sf n} \sim 100$, the corresponding $\overline{A} \sim 0.1$. As we will soon outline in discussing dynamics
(Section \ref{sec:dynamics}), such a value of $\overline{A}$ is that typically required to fit the viscosity and dielectric response measurements. In general, with the above approximations in tow,
\begin{eqnarray}
\label{nd}
{\sf n} \approx \frac{1}{3 \overline{A}^2}.
\end{eqnarray}
In these estimates, we used the classical Dulong-Petit heat capacity of an 
equilibrium harmonic solid. In general (yet not always), the classical heat capacity
of fluids interpolates between that of  the Dulong-Petit form of harmonic solids (that we 
used above to obtain estimates) and that of gases at yet higher temperatures (being half the Dulong-Petit value). Fluids do not support shear and thus their transverse modes contribute less to the heat capacity than phonons do in a solid \cite{Kostya-elastic}.
Similarly, as the temperature is depressed in the equilibrium  solid, the heat capacity is lower than its classical Dulong-Petit value due to quantum effects. Thus, the Dulong-Petit value that we employed in the above estimates is an upper bound on the heat capacity. This will lead to slightly lower estimates of $n$ given a value of $\overline{A}$. 

Given Eq. (\ref{nd}), for $0.05 \lesssim \overline{A} \lesssim 0.15$ (values that viscosity measurements suggest for an entire spectrum of fluids), $135 \gtrsim {\sf n} \gtrsim 15$ or, equivalently, a cube of linear dimension of $~2.5-5~$ atoms on a side. Such a size is qualitatively consistent with the dimensions of spatial structures reported in supercooled liquids \cite{3D_Structure} via atomic electron tomography reconstruction as well as dynamical heterogeneity length scales (growing with decreasing temperatures) observed in correlation electron microscopy experiments \cite{volynes}. A priori, the effective ``${\sf n}$'' may change with temperature yet setting it to a constant yields excellent viscosity fits. This is the scale over which the system effectively manages to locally equilibrate in maximizing its entropy.

The standard deviation of the  distribution $P(T|T')$ must, by dimensional analysis, be of the form 
\begin{eqnarray}
\sigma_T = T   F (\frac{T}{T_{\sf melting}},~ \cdots ),
\end{eqnarray}
where $F$ is a general function and the ellipsis denote dimensionless quantities (including all ratios of energy or temperature scales specific to a particular supercooled fluid such as the system freezing/melting temperature). Physically (since at zero temperature, the energy density of the supercooled system cannot be lower than that of the annealed crystalline system) and as is indeed captured by the above scaling form, we anticipate that as the temperature $T \to 0$, in the absence of relevant $T'<0$, the standard deviation $\sigma_T$ of an assumed Gaussian $P(T|T')$ must vanish. At temperatures far away from the system melting and all other special scales, $F$ may tend to the constant value (${\overline{A}}$). Sufficiently high cooling rates at low temperatures may thus render void a description with a static $\rho_{\sf eff.}$ Our focus is on the situations in which the supercooled fluid is still in effective equilibrium. The distribution $P$ might, of course, be bimodal or other with additional nontrivial scales and more complicated yet contributions. Indeed, as underscored in Eq. (\ref{superposition}), $P(T|T')$ may be a (non-unique) mixture of such Gaussians. For hydrodynamic flow, the support of the distribution from the equilibrium fluid contributions (assumed to be the above normal distribution) from higher energy densities $\epsilon'$ (or temperatures $T'$) might be the most important. This may arise from higher values of $c_\ell T$ in Eq. (\ref{superposition}) or low cluster sizes $n$ (for which the standard deviation $\overline{A}_{\ell}$ T may become large). For simplicity, we will assume the form of Eq. (\ref{cluster}) throughout.

\subsection{A relation between configurational heat capacity and energy broadening}
\label{sec:config}
  To plainly demonstrate a basic qualitative connection, in this subsection we will continue to consider  the schematic suggested in Section \ref{hydro} of local entropy maximization for ${\sf n}-$particle clusters. In a simple, error-analysis type, approximation the standard deviation of the energy (stemming from the distribution $P(T|T')$) is uncorrelated with and will  approximately add in quadrature with the standard deviation of the equilibrium energy. In such a situation, for $(N/{\sf n})$ independent ${\sf n}-$particle clusters,
\begin{eqnarray}
\label{quad}
\frac{\sigma^2_E,s.c.}{(N/{\sf n})} =\frac{\sigma_{E.eq.}^2}{(N/{\sf n})} + \sigma_{\epsilon}^2.
\end{eqnarray}
 In Eq. (\ref{quad}), $\sigma_{\epsilon}$ 
denotes the excess standard deviation of the energy $\epsilon_{s.c.}$ of the ${\sf n}-$particle cluster over its equilibrium counterpart; $\sigma_{\epsilon}$ is non-zero since the conditional probability $P$ is not a delta function in the energy per cluster $\epsilon$. In the canonical setting, both the equilibrium energies of ${\sf n}-$particle clusters as well as the energy variances are linear in the particle number. Thus, for different sizes ${\sf n}$, the dimensionless constant $\overline{A}$ may scale with ${\sf n}$ (as was the case for Eq. (\ref{localcv})). 
As we will demonstrate, 
\begin{eqnarray}
\label{energy_var_sc}
\sigma^2_{E,s.c.} = k_{B} T^2 C_{v,s.c.},
\end{eqnarray}
with $C_{v.s.c.}$ being the constant volume heat capacity of the supercooled fluid. The equilibrium energy fluctuations $\sigma_{E,eq}^2 = k_{B}T^2 C_{v,eq.}$. 

Eq. (\ref{quad}) implies that 
\begin{eqnarray}
\label{set}
\sigma_{\epsilon}^2 = k_{B} T^2 \Delta c_{v},
\end{eqnarray}
where $\Delta c_{v}$ 
is the difference between the constant volume specific heat of the ${\sf n}-$particle cluster in the supercooled fluid 
and ${\sf n}-$particles in the equilibrium crystal, 
\begin{eqnarray}
\label{deltacv}
\Delta c_{v}
\equiv \frac{C_{v,sc} - C_{v,eq}}{(N/{\sf n})}.
\end{eqnarray}
The heat capacity difference $(C_{sc} - C_{eq})$ has been studied extensively over the years.
In particular, integration of the experimentally measured heat capacity $(C_{sc} - C_{eq})/T$ over the temperature $T$ (experimentally, the latter are measured at constant pressure not volume) yields the entropy difference between the supercooled liquid and the crystal- the so called ``configurational entropy'' whose extrapolation underlies the famous Kauzmann paradox \cite{Walter} which we will briefly return to in Section \ref{sec:thermo}. In Eq. (\ref{deltacv}), the cluster heat capacity difference $\Delta c_{v}$ 
is ${\sf n}$ times the standard measured (i.e., that per particle) excess specific heat of the supercooled liquid as compared to that of the equilibrium crystal. 

We next further justify and elaborate on the appearance of the supercooled heat capacity in Eq. (\ref{set}) when estimating the variance of the energy of the supercooled fluid. Towards this end, we note that (applying Eq. (\ref{pp}) to a Hamiltonian that is a sum of local few body terms), the average energy per cluster in the supercooled fluid,
\begin{eqnarray}
\label{localE}
\epsilon_{\sf s.c.}(\beta) = \int d \beta' P(\beta|\beta') \epsilon_{\sf eq.}(\beta').
\end{eqnarray}
We stress that here an assumption has been made that the energy of the supercooled fluid can be expressed as the sum of the corresponding appropriate local ${\sf n}-$particle energy densities. There is, therefore, a minimal value of ${\sf n}$ for which Eq. (\ref{pp}) is applied (set by the range of the interactions). This minimal value of ${\sf n}$ is the one for which our below derived estimates may apply. 
If $P(\beta|\beta')$ is predominantly a function of $(\beta - \beta')$ (as in, e.g., the conditional probability of Eq. (\ref{cluster}) for small $\overline{A}$) then noting that $\partial_{\beta} P = - \partial_{\beta'} P$, we  explicitly have
\begin{eqnarray}
\label{longcv}
- \frac{\partial \epsilon_{\sf s.c.}}{\partial \beta} \nonumber
\\ = - \int d \beta' \epsilon_{\sf eq.}(\beta') \frac{\partial P(\beta- \beta')}{\partial \beta} \nonumber
\\ = \int d \beta' \epsilon_{\sf eq.}(\beta') \frac{\partial P(\beta- \beta')}{\partial \beta'} \nonumber
\\ =
 \int d \beta' (-\frac{\partial \epsilon_{\sf eq.}(\beta')}{\partial \beta'})  P (\beta- \beta') \nonumber
\\ =  \int d \beta' \langle (\Delta \epsilon)^2 \rangle_{\sf eq., \beta'} ~ P(\beta'|\beta)  \nonumber
\\ = \langle (\Delta \epsilon)^2 \rangle_{\sf s.c.}.
\end{eqnarray}
Here, $\langle (\Delta \epsilon)^2 \rangle_{\sf eq.}$ and $\langle (\Delta \epsilon)^2 \rangle_{\sf s.c.}$ refer, respectively, to the energy variance of the ${\sf n}$ particle cluster in equilibrium and in the supercooled fluid. As seen from Eq. (\ref{longcv}), similar to standard equilibrium thermodynamics, the variance of the energy in the supercooled fluid is given by the derivative of its energy relative to the inverse temperature. Thus whenever $P(\beta|\beta')$ is a function of $(\beta- \beta')$, we may indeed replace, in Eq. (\ref{quad}), the scaled variance of the energy density (i.e., the energy per cluster)  $\frac{\sigma^2_{E, {\sf s.c.}}}{N/n}$ by $(- \frac{\partial \epsilon_{\sf s.c.}}{\partial \beta} ) = k_{B} T^2 c_{v,{\sf s.c.}}$, establishing Eq. (\ref{energy_var_sc}). For completeness, we briefly elaborate on all of the simple steps in the derivation of this relation. In the third equality of Eq. (\ref{longcv}), we integrated by parts and in fourth equality, we used the fact that the derivative of the equilibrium energy density $(-\frac{\partial \epsilon_{\sf eq.}(\beta')}{\partial \beta'})$ is the variance of the equilibrium energy at an inverse temperature $\beta'$. In the last equality of Eq. (\ref{longcv}), we employed Eq. (\ref{pp}). This substitution then yields Eq. (\ref{energy_var_sc}) and hence establishes Eq. (\ref{set}). For  $P(\beta|\beta')$ that are not solely a function $(\beta - \beta')$, there will be additional corrections to Eq. (\ref{longcv}) which may become sizable (e.g., for the distribution of Eq. (\ref{cluster}) for sufficiently large  $\overline{A}$). As we noted earlier, what is experimentally measured is the excess of the constant pressure specific
heat 
(not that of the specific heat at constant volume). 

To obtain order of magnitude estimates, we next consider what will transpire if the heat capacity difference $\Delta c_v$ is approximately constant in the studied temperature range above from the glass transition temperature
($T_{melting} \ge T > T_g$). In such a case, since from Eq. (\ref{cluster}), the standard  deviation of the effective equilibrium temperature $T'$ as computed with $P(T|T')$ is $\sigma_{T} = \overline{A} T$. Because the standard deviation of the energy scales as the average heat capacity of the $n-$particle cluster ($\overline{c}$) times the latter standard deviation of the temperature, i.e., $\sigma_{\epsilon} \sim \overline{c} \sigma_{T} \sim \overline{c} \overline{A} T$. 
Eq. (\ref{set}) then implies that 
\begin{eqnarray}
\label{AA}
\overline{A} \sim \sqrt{\frac{k_{B} \Delta c_v}{\overline{c}^2}}.
\end{eqnarray}
If $\Delta c_v \sim \overline{c} \sim 3 {\sf n} k_{B}$ then 
we will reobtain Eq. (\ref{nd}). 
What is measured in experiment is the specific heat (that per particle) which is a factor of ${\sf n}$ (the effective cluster size) smaller than the above quantities. Empirically, 
the constant volume specific heat of the equilibrium system (the crystal) is indeed of the same scale as the specific heat difference between the equilibrium crystal and the supercooled fluid, e.g., \cite{PVAC} (and of similar behavior to that of the constant pressure heat capacity \cite{Tang}).

Although the relative energy density fluctuations (for an $n_a$ particle subsystem of ${\sf n}$) may typically increase as the subsystem becomes smaller (i.e., the standard deviation of the energy of $n_a$ particles divided by their total energy), the effective temperature broadening as seen in $P(\beta|\beta')$ appears for all reduced local probability densities of $n_a \le {\sf n}$ particles. 
What matters for the simple estimates in the current subsection and Section \ref{hydro} 
is that local equilibrium is established and the system appears stationary for observables on the $n={\sf n}-$particle assemblies (whether these might form a cluster or not). The size of these assemblies ${\sf n}$ (or the maximal number of particles $n$ for which local stationary is seen (and thus Eq. (\ref{pp}) holds) determines the all too simple scaling in Eqs. (\ref{nd},\ref{AA}).
For the reason discussed following Eq. (\ref{localE}), the particle number ${\sf n}$ cannot be smaller than the number of particles that lie within the range of the interactions.

\subsection{Dynamics}
\label{sec:dynamics}
 
Augmenting the single particle velocity  distributions within the supercooled liquid (Section \ref{sec:extract}), we may examine quantities that may be ascertained by conventional experimental  measurements such as those relating to the shear viscosity. Our framework employs local reduced probability densities. Thus, regrettably, we cannot apply the Green-Kubo equation to directly  compute the shear viscosity (the latter involving stress correlations over the entire system). Towards that end, we will use another (gedanken experiment) device and compute the viscosity as it is often measured in experiment by a local measurement- that of inserting a sphere and measuring its terminal speed. Such an approach for predicting the viscosity was employed in Refs. \cite{longrange,PS,1parameter,critical} although there the underlying analysis focused on the global probability density (as opposed to the local reduced probability density of the current work).  

As we emphasized in the earlier Sections, although $\rho_{\Lambda}$ evolves with time, the macrostate of the system as defined by local few body measurable quantities is effectively time independent at times $t_{\min} \le t' \le t_{\sf xtal}$. 
That is, we may compute long time average within the steady state of the  supercooled liquid with the aid of Eq. (\ref{pp}) and its extensions (e.g., Eq. (\ref{pp'})). Complementing the single liquid particle distributions of Section \ref{sec:extract}, we may also compute the velocity distribution of inserted test particles that exert shear. In particular, we can supplement the Hamiltonian of the liquid by a gravitational term $(- m_{\sf sphere} g z_{\sf sphere})$, with $g$ the  acceleration of free fall, to captures the addition of a dropped sphere (radius $R$) of mass $m_{\sf sphere} = \frac{4 \pi}{3} \rho_{\sf sphere} R^{3}$ from a height $z_{\sf sphere}$ into the supercooled liquid. If the mass density of the supercooled liquid is $\rho_{\sf liquid}$ then, from the Stokes relation
applied to the high viscosity (low Reynolds number) fluid, the terminal (i.e., long time average of the) velocity of the dropped test sphere is
\begin{eqnarray}
\label{stokes}
    v_{\sf s.c.} = \frac{2}{9} \frac{\rho_{\sf sphere} - \rho_{\sf fluid}}\eta gR^2.
\end{eqnarray}
Such a long time average is that associated with the reduced probability density of Eq. (\ref{rlta}) for which we established Eq. (\ref{pp}). 
The Stokes relation of Eq. (\ref{stokes}) allows us to compute the  shear viscosity $\eta$ by calculating the average of the speed of the sphere in the supercooled liquid $v_{\sf s.c.}$ with its reduced probability density (that may include a finite number $n$ of other fluid particles) and then inverting Eq. (\ref{stokes}).  
Indeed, once the falling sphere and supercooled liquid reach a steady state, we may use Eq. (\ref{pp}) in its simplest (i.e., $n=1$) most local form  
to relate a single particle reduced probability density-
that of the falling sphere- in both the supercooled fluid and when it dropped and a steady state is achieved in the bona fide equilibrium liquid or solid. Specifically, for a supercooled fluid system of energy density $\epsilon$ at temperature $T$,
\begin{eqnarray}
\label{vsc}
v_{{\sf s.c.}}= \int dT' P(T|T') v_{\sf eq.}(T') \nonumber
\\ + \int_{\cal PTEI} d \epsilon' P(\epsilon|\epsilon') v_{\sf eq.} (\epsilon').
\end{eqnarray}

\begin{figure*}
	\centering
	\includegraphics[width=2 \columnwidth, height=.7 \textheight, keepaspectratio]{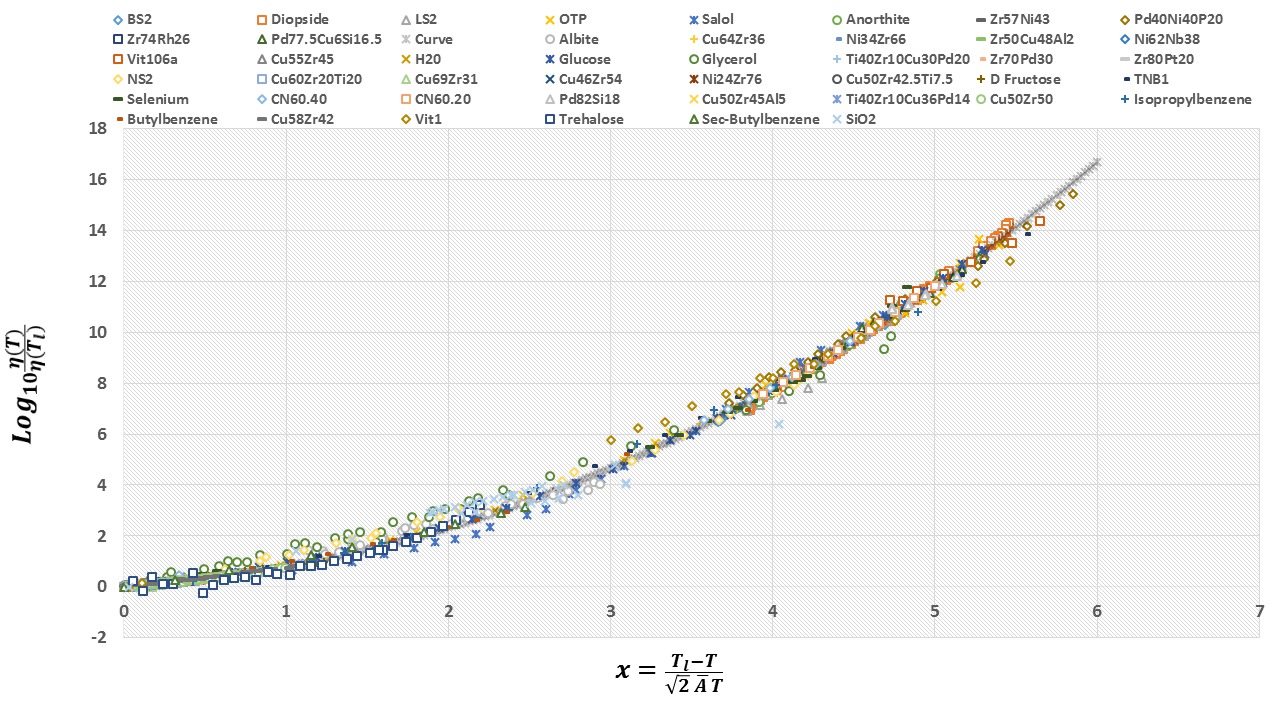}
	\caption{(Color Online.) 
 Reproduced from \cite{PS}. A test of  Eq. (\ref{eq:visc1}) for 45 supercooled liquids. Along the vertical axis, the viscosity data are scaled by their value at the melting (or liquidus) temperature, $\eta(T_{l})$ to form dimensionless quantities. The horizontal axis provides the 
 (scaled and shifted) temperature in dimensionless units $x \equiv \frac{T_{l} - T}{\sqrt{2} \overline{A} T}$. The viscosity data of all 45 different liquids from numerous classes/bonding types (silicate, metallic, organic) and kinetic fragilities collapse onto a unique curve (fitted above by Eq. (\ref{eq:visc1})), suggestive of universality amongst all types of glassforming liquids. Note the exceptional agreement over 16 decades. In contrast to other theories \cite{Cavagna,BB}, {\it no additional conjectured temperatures appear} beyond the experimentally measured equilibrium melting (or liquidus) temperature.} 
	\label{Collapse.}
\end{figure*}

When the applied external stress (excess weight of the test sphere in Eq. (\ref{stokes}) relative to that of the surrounding medium into which it is dropped) tends to zero \footnote{Crystalline defects (e.g.,  dislocations) may drift following the application of a finite external stress. Our focus is, however, on infinitesimal external stress following the definition of the shear viscosity of the equilibrium system as given by linear response in that limit.}, the viscosity of equilibrium crystalline solids diverges \cite{sausset}. Thus, assuming that equilibrium liquid flow occurs only above the equilibrium melting (or liquidus) temperature (or, more precisely, the energy density at the upper end of the ${\cal PTEI}$ region) and substituting Eqs. (\ref{cluster},\ref{vsc}) in Eq. (\ref{stokes}), 
\begin{eqnarray}
\label{eq:visc1}
\eta_{\sf s.c.}(T) \simeq \frac{\eta_{\sf eq.}(T^+_{\sf melt})}{\int_{T_{\sf melt}}^{\infty} dT' P(T|T')} =  \frac{\eta_{\sf s.c.}(T^+_{\sf melt})}{{\sf erfc}  (\frac{T_{\sf melt} - T}{\overline{A}T \sqrt{2}})},
\end{eqnarray}
with ${\sf erfc}(x)$ the complimentary error function. The final (second)  equality in Eq. (\ref{eq:visc1}) is more specific than the first one in assuming that the conditional probability $P$ is of the scale free Gaussian form of Eq. (\ref{cluster}). Eq. (\ref{eq:visc1}) was derived and studied in \cite{1parameter,PS,longrange,local_structure,critical}. In these works, the standard deviation of the energy density of the entire system was taken to be finite as opposed to the analysis of the reduced local probability densities that we study here (for which the system does not have a finite standard deviation of its energy density). Empirically, Eq. (\ref{eq:visc1}) fits the viscosities of all supercooled liquids rather well (see Figure \ref{Collapse.}) with the aforementioned constant $\overline{A}$ (whose precise value (i.e., the one optimal for fitting the experimental viscosity data) varies from one supercooled liquid to another) being, typically, $\overline{A} \sim 0.1$ \cite{longrange,PS}.

Similarly, using the approximation of Eqs. (\ref{Gauss},\ref{quad},\ref{set}) instead of Eq. (\ref{cluster}), 
\begin{eqnarray}
\eta_{\sf s.c.}(T) \simeq  \frac{\eta_{\sf s.c.}(T^+_{\sf melt})}{{\sf erfc}  (\frac{\overline{c}(T_{\sf melt} - T)}{\sigma_{\epsilon} \sqrt{2}})}
= \frac{\eta_{\sf s.c.}(T^+_{\sf melt})}{{\sf erfc}  (\frac{\overline{c}(T_{\sf melt} - T)}{\sqrt{2k_{B}T^{2} \Delta c_v}})},
\end{eqnarray}
with the average specific heat $\overline{c}$ of Eq. (\ref{av:c}). Far more directly, refraining from such approximations, we can compute the viscosity as
\begin{eqnarray}
\label{viscosityfromv}
\eta_{\sf s.c.}(T) \simeq \frac{\eta_{\sf eq.}(T^+_{\sf melt})}{\int_{T_{\sf melt}}^{\infty} dT' P(T|T')}.  
\end{eqnarray}
That is, by employing Eq.(\ref{Pfromv}), we may infer $P(T|T')$ from the velocity distribution that is observed in simulation/experiment. We may then insert this computed $P(T|T')$ into Eq. (\ref{viscosityfromv}) to obtain a prediction for the viscosity that is free of assumptions of what the distribution $P(T|T')$ looks like. The relaxation dynamics that we derive self consistently relate to the change of the local reduced probability density and thus implicitly to the change of the effectively nearly constant ``adiabatic'' conditional probability distribution $P$ itself (Section \ref{finite_time}). The latter changes of the probability densities will be of ever smaller magnitude at low temperatures where the integral in the denominator of Eq. (\ref{eq:visc1}) is vanishingly small and thus (as is  consistent with experiment (Figure \ref{Collapse.})) the viscosity and associated structural relaxation (and crystallization) times are extremely large. The higher moments of the long time average velocity distribution of the dropped test spherical particle that led to the prediction of Eq. (\ref{eq:visc1}) may be readily computed. In the steady state Galilean reference frame of the dropped sphere, the long time velocity average of the sphere is (by definition) zero as are those of all of its odd moments (consistent with the trivially anticipated cancellations of the velocity  fluctuations in the long time average). The even moments of the velocity relative to the center of mass of the sphere are given by Eq. (\ref{v-equil-moment})
(or, more precisely, by evaluating these  moments with the distribution of Eq. (\ref{no_Maxwell_dist}) when no assumptions are made regarding the form of the conditional probability distribution $P$). Whenever measurable, these velocity moments of the dropped test sphere (as well as those of the particles of the supercooled liquid) may be contrasted with those of the viscosity to see whether the same conditional probability distribution $P$ consistently leads to all of these.   \\

The evolution of the reduced (unlike that of the full system governed by a Liouvillian ${\hat{\cal{L}}}$ \footnote{The full many body probability density matrix evolves as 
\begin{eqnarray}
\rho_{\Lambda} (t') = e^{-i  {\hat{\cal{L}}} t'} \rho_{\Lambda}(0).
\end{eqnarray}
Classically, the Liouvillian is given by
\begin{eqnarray}
{\hat{\cal{L}}} \cdot  = i \{ H, \cdot \}_{P.B.}
\end{eqnarray}
with $\{ , \}_{P.B.}$ denoting a Poisson bracket. Quantum mechanically,
\begin{eqnarray}
    {\hat{\cal{L}}} \cdot =  \frac{1}{\hbar} [H, \cdot]. 
\end{eqnarray}
}) probability density is, in general, not unitary. Nonetheless, \\

{\bf (A)} The reduced local probability densities of the supercooled and equilibrium systems are related by a {\it homogeneous linear equation}
(Eq. (\ref{pp})). \newline

Similarly,  \newline

{\bf (B)} The equations of motion for these finite order reduced probability densities (e.g., the classical BBGKY hierachy of Eqs. (\ref{BBGKY-eq}) or the  quantum Nakajima?Zwanzig \cite{NZ1,NZ2} or Lindblad \cite{Lindblad} equations)  are {\it homogeneous linear equations}. \\

This homogeneous linear character of the equations of motion in the finite ($n$-th) order reduced local probability densities implies that all time dependent observables averaged with an initial probability density in the supercooled system may remain linearly related (via Eq. (\ref{pp})) to those in the equilibrium system. When the equations of motion are integrated forward in time, the evolution of any particle will, in general, involve contributions from an ever increasing number of particles that influence it directly or indirectly (i.e., by interacting with other particles that do) as is captured in the BBGKY hierarchy (briefly reviewed in the Appendix, Eqs. (\ref{BBGKY-eq})). For sufficiently high orders (i.e., large enough particle number $n$) the observed reduced probability density is no longer stationary and consequently Eq.(\ref{pp}) might break down. In particular, the BBGKY hierarchy for the reduced probability densities in the supercooled liquid can then involve reduced probability densities that are not related by Eq. (\ref{pp}) to their counterparts in the equilibrium system. Relations such as those that we motivate next then become void. Given, however, that for small $n$, the reduced probability densities $\rho_{n}$ appear stationary for long times in the supercooled fluid, it is possible that the time scale over which these higher order corrections become important may be very large (of the order of the $\tau_{\sf xtal}$). So long as the decomposition of Eq. (\ref{pp'}) holds,  
various dynamical metrics of the supercooled liquid can be related to those in the equilibrium system governed by the same Hamiltonian. For instance, the above suggests that the average squared displacement of a given particle  is 
\begin{eqnarray}
\label{displacement}
&& \langle ({\bf \Delta  x}_{i}(t))^2 \rangle_{{\sf s.c.~},T,\mu, \cdots} \nonumber
\\
 &&  = \frac{1}{N}{\sf Tr}\Big(\rho_{\sf s.c.}~  \Big( {\bf \Delta x}_{i}(t) \Big)^2 \Big) \nonumber
 \\ && =
\int d\beta' d \mu'  \ldots \Big(P(\beta,\mu, \ldots| \beta',\mu
',\ldots) \nonumber
\\ && \times
\langle ({\bf \Delta x}_{i}(t))^2 \rangle_{{\sf eq.~},\beta',\mu', \cdots} \Big) \nonumber
\\ && + \int_{\cal{PTEI}} d\epsilon'd {\tilde{\sl{n}}}' \cdots 
\Big(P(\epsilon, {\tilde{\sl{n}}}, \ldots| \epsilon',{\tilde{\sl n}}
',\ldots)  \nonumber
\\ && \times \langle ({\bf \Delta x}_{i}(t))^2 \rangle_{\sf eq., \epsilon', {\tilde{\sl n}}', \cdots} \Big).
\end{eqnarray}

The time dependent expectation value $\langle ({\bf\Delta x}_{i}(t))^2 \rangle_{{\sf eq.~},\beta',\mu', \cdots}$ of the equilibrium system evolves as determined by the Liouvillian ${\hat{\cal{L}}}$. Given the character of the different equilibrium phases, one may then anticipate (as is indeed qualitatively consistent with observations  \cite{BB}) that intermediate time caging like behavior will arise from the equilibrium ``solid'' phonon like modes from high 
$\beta'$ (i.e., from the low temperature equilibrium solid)
and long time diffusive contributions will be borne from 
``fluid'' like modes of low 
$\beta'$ (high temperature equilibrium fluid). Phonon type contributions to the average squared displacement $\langle ({\bf \Delta  x}_{i}(t))^2 \rangle_{{\sf s.c.~},T,\mu, \cdots}$ may, as in the equilibrium solid, be bounded in size and saturate to a fixed constant value at large times $t$. These equilibrium crystalline and equilibrium liquid contributions to relaxation dynamics are, respectively, reminiscent of the short time  ``beta'' and much longer time ``alpha'' type  relaxation processes in supercooled  liquids \cite{Cavagna,BB} that in turn may be further cast as known distributions (see \cite{Bello} for a review) of simple relaxation processes.  Disparate equilibrium solid and fluid like contributions  naturally stem from a distribution of local structures and  associated elastic stresses. We will further touch on these in Section \ref{sec:structure}.    

We can approximate, in an analogous manner, the self part of the van Hove function \cite{BB}. This function provides the probability that the displacement a particle in the supercooled fluid over a time interval $t$ will be equal to ${\bf x}$, 
\begin{eqnarray}
\label{GVH}
 G_{vH, {\sf s.c.}}({\bf{x}},t) \equiv \langle \delta({\bf{x}} - {\bf{x}}(0) + {\bf{x}}(t)) \rangle_{{\sf s.c.~},T,\mu, \cdots}.   
\end{eqnarray}
Similar to Eq. (\ref{displacement}), the self part of the van Hove function for a particle in the supercooled fluid at temperature $T
$, chemical potential $\mu$,  $\cdots$, may be approximated by a weighted average of the self part of the van Hove function of the equilibrium system at different equilibrium temperatures $T'$, chemical potentials $\mu'$, etc. The self part of the van Hove function of Eq. (\ref{GVH}) is often used as a metric that quantifies dynamical heterogeneity \cite{BB}. For fixed time $t$, the function $G_{vH,{\sf s.c.}}({\bf{x}},t)$ exhibits long distance Gaussian tails (that may be associated with fluid-like motion) with short distance (small $|{\bf{x}}|$) narrow peak about the origin (that can be interpreted as a solid-like contribution). For an equilibrated system at temperature $T'$ with a diffusion constant $D_{\sf eq.}(T')$, in $d$ spatial dimensions, at long times $t$, the self part of the van Hove function is given by the Gaussian $\frac{1}{\sqrt{4 d \pi D_{\sf eq.}(T') t}} e^{-(\Delta {\bf{x}}(t))^{2}/(2dD_{\sf eq.}(T') t)}$.  Thus, at long times $t$, for a supercooled fluid of temperature $T$, 
\begin{eqnarray}
G_{vH,{\sf s.c.}}(t) \sim  \int dT' ~P(T|T')~ \frac{1}{\sqrt{4 d \pi D_{\sf eq.} (T') t}} \nonumber
\\ \times e^{-(\Delta {\bf{x}}(t))^{2}/(2dD_{\sf eq.} (T')t)}.
\end{eqnarray}
The same superposition that we are led to with $\rho_{\sf eff}$ having both solid-like and fluid-like contributions 
(with $T'$ larger than or smaller than the equilibrium melting temperatures) applies for general (both linear and higher order) susceptibilities. These  susceptibilities will be smeared averages of their values in the equilibrium crystalline and fluid phases 
and their coexistence in the ${\cal PTEI}$. For instance, the linear finite frequency susceptibilities,
\begin{eqnarray}
  && \chi_T(\omega)  \nonumber
  \\ && = \int d T' ~ d \mu' \cdots   \Big(P(T,\mu, \cdots|T',\mu', \cdots) \nonumber
 \\  && \times
\chi_{\sf eq.}(\omega')_{T',\mu',\cdots} \Big) \nonumber
\\
 && + \int_{\cal{PTEI}} d \epsilon' d n' \cdots  \Big( P(\epsilon,n, \cdots|\epsilon',n', \cdots) \nonumber
\\  
 && \times \chi_{\sf eq.}(\omega')_{\epsilon',n',\cdots}\Big)~,  
\end{eqnarray}
will have contributions from different equilibrium phases.
As will be further detailed elsewhere, nearly identical considerations also apply to the intermediate scattering function and other measures \cite{Nick-to-be}. 

Similar to the discussion in Section \ref{consequences}, a key point worth emphasizing is that the very same conditional probability $P$ yields, in unison, predictions for all of the dynamical measurements discussed here such as the viscosity, single particle displacement, self part of the van Hove function, linear finite frequency 
as well as general higher order susceptibilities and the thermodynamic and structural properties that we will shortly turn to. Some other data not shown here (e.g., dielectric response times \cite{Nick-to-be}) were found to be consistent with these predictions.

\subsection{Thermodynamic measurements}
\label{sec:thermo}

\begin{figure*}
\includegraphics[width=2 \columnwidth, height=.75 \textheight, keepaspectratio]{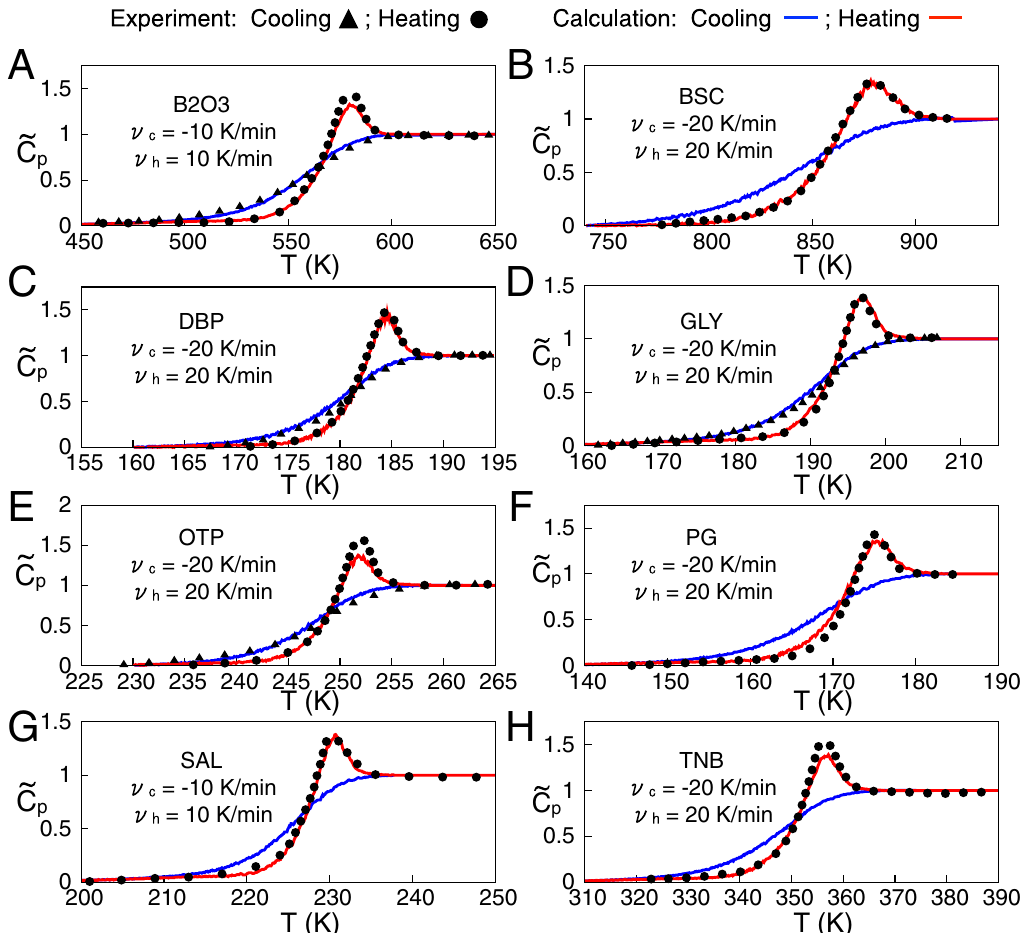}
\caption{(Color Online). Reproduced from \cite{Keys}. Scaled constant pressure specific heat as a function of temperature on cooling and heating. At high temperatures, the scaled specific heat tends to unity. Real experimental data points for these fluids are plotted on top of the curves suggested by \cite{Keys}. As long known, close to the dynamic glass transition temperature $T_g$ (at which the viscosity is $10^{12}$ Pascal $\times$  sec.), a peak in the specific heat is seen on heating (red curve). This defines the ``calorimetric glass transition temperature'' which is often nearly indistinguishable from the dynamic glass transition temperature. 
The latter peak is absent on cooling (blue). Where viscosity data are available, this enables us to test the estimate of Eq. (\ref{scale-c-peak}) (see also \cite{longrange}) 
For OTP, $T_g = 240$ K while the separation between the heating and cooling curves persists over a temperature window $\Delta T_{h} \approx 20$ K (between temperatures of 235 and 255 K). Given that the empirical value of $\overline{A}$ required to fit the OTP viscosity data with Eq. (\ref{eq:visc1}) is $\overline{A} = 0.049$ \cite{PS}, the ratio $\Delta T_{h}/(2\overline{A} T_{g}) \approx 0.85$. Similarly, for Glycerol (GLY), having a glass transition temperature of $T_{g}= 190$ K, and a value of $\overline{A} = 0.077$ \cite{PS} from its viscosity data \cite{PS}, the heating and cooling curves differ between 175 and 200 K ($\Delta T_h =25$ K). Thus, for glycerol, $\Delta T_{h}/(2\overline{A} T_{g}) \approx 0.85$. For Salol and TNB, only theoretical ``computed" data appear for cooling.
Assuming the data and analysis of \cite{Keys} to be correct and thus taking for Salol
$\Delta T_h \approx 25~K$, given that $T_g = 220~K$ and, from the viscosity fit, $\overline{A} = 0.062$ \cite{PS}, 
the ratio $\Delta T_h/(2\overline{A} T_g) \approx 0.92$. For TNB, from the above figure, $\Delta T_{h} \approx 40~K$, the viscosity fit $\overline{A} = 0.054$ \cite{PS}, and $T_{g} = 345~K$ \cite{Richert} leading to $\Delta T_h/(2\overline{A} T_g) \approx 1.07$. From Eq. (\ref{scale-c-peak}), we anticipate the ratio $\Delta T_h/(2\overline{A} T_g) \approx 1$ which is qualitatively consistent for the above three fluids for which viscosity data enabled the determination of $\overline{A}$. For the other four fluids (B$_2$O$_3$, BCS, DBP, and PG) there were no viscosity data in \cite{PS} that enabled the extraction of the parameter $\overline{A}$ and thus the evaluation of the above ratio.}
\label{Specific_heat.}
\end{figure*}

Predictions for thermodynamic quantities may be performed along lines similar to our analysis for the dynamics. Towards this end, we set, in Eq. (\ref{Qavglong}), $Q$ to be an intensive thermodynamic state function that can be expressed as the sum of local operators (for which the average with $\rho_{\sf eff.}$ is well defined, e.g., the energy density). Doing so allows us to relate 
the thermodynamic state function densities of the supercooled system to those of the equilibrium crystal or fluid \cite{1parameter}. Since the equilibrium system has a region of phase coexistence, a consequence of Eq. (\ref{pp'}) is that the state (and other) functions of the supercooled system 
may exhibit two crossovers at the temperatures associated
with the two limiting values of the energy densities between which the ${\cal PTEI}$ of the latter equilibrium (i.e., non supercooled) system exists. This corollary is consistent with empirical observations and various theoretical perspectives of a putative ``liquid-liquid'' transition \cite{liquid-liquid} below the liquidus temperature in supercooled liquids. Whether there is any relation \cite{PS} to liquid-liquid transitions can only be a speculation at this stage. These and other effects may, however, be obscured by the implicit finite time (frequency) of the thermodynamic probes. Since the supercooled system ``falls out of equilibrium'' 
on experimental time scales below the ``glass transition'' temperature $T_g$ where the relaxation time becomes so long such that all measurements are, in practice, effectively performed at a finite cutoff frequency. Indeed, if the time scale for the system relaxation is much shorter than the effective set cutoff then this is not an issue. However, at sufficiently low temperatures, the measurement might capture the values of the bona fide thermodynamic (and other) quantities. As systems fall out of equilibrium, memory of their history may appear. This is often manifest in hysteresis curves.
In supercooled fluids, on the relevant experiment time scales, a hysteresis of the specific heat is indeed observed near the glass transition temperature, see, e.g., Figure \ref{Specific_heat.}. Such a hysteresis may be a dynamical crossover effect associated with the above noted finite frequency cutoff for the specific heat
(and other) measurements. That is, the system cannot attain a nearly stationary metastable state on the measured time scale and, on cooling, still retains memory of its history at higher temperatures. The low temperature system being heated similarly ``recalls'' its earlier rigid glassy state and exhibits a heat capacity peak somewhat reminiscent to that of melting. 
 In other words, given the distribution of Eq. (\ref{cluster}), one may anticipate that as the system is heated or cooled it will start to ``sense'' the dynamic crossover at $T_g$ at temperatures proximate to $T_g \pm \sigma_{T_g} = T_{g} (1 \pm \overline{A})$. One may hence suspect \cite{longrange} that the width of the hysteresis of the measured heat capacity (Figure \ref{Specific_heat.}) may scale as 
\begin{eqnarray}
\label{scale-c-peak}
    \Delta T_h \simeq 2 \sigma_{T_{g}} \simeq 2\overline{A} T_g.
\end{eqnarray}

This qualitative prediction may be tested. As briefly noted in \cite{longrange}, we may contrast such a relation against experimental measurements. 
In Ref. \cite{Keys}, the heat capacities of several fluids were examined. Apart from OTP, Glycerol, Salol, and TNB, none of the other four fluids appearing in Figure \ref{Specific_heat.} are those for which viscosity data led to an extraction of $\overline{A}$ in \cite{PS} enabling a calculation of the above ratio. For OTP, Glycerol, Salol, and OTP, with data from \cite{Keys} (see Figure \ref{Specific_heat.}), Eq. (\ref{scale-c-peak}) is qualitatively obeyed. Detailed analysis for other fluids will be reported elsewhere \cite{Jing}. Using Eq. (\ref{eq:visc1}), the  relation of Eq. (\ref{scale-c-peak}) may be further cast in a form that does not include $\overline{A}$ nor any other fitting parameter, 
\begin{eqnarray}
\label{scale=1}
\Delta T_{h} \simeq \frac{(T_{\sf melt} - T_{g}) \sqrt{2}}{{\sf erfc}^{-1}\Big(\frac{\eta_{\sf s.c.}(T_{g})}{\eta_{{\sf s.c.}}(T^{+}_{\sf melt})}\Big)}.
\end{eqnarray}
By definition, at the dynamical glass transition temperature, the shear  viscosity assumes a fixed value \cite{Cavagna,BB} ($\eta_{\sf s.c.}(T_{g}) = 10^{12}$ Pa $\times$ sec.) thus leaving our estimate of Eq. (\ref{scale=1}) to depend only on the value of the equilibrium melting (or liquidus)  temperature $T_{\sf melt}$, the viscosity of the supercooled fluid just above melting $\eta_{\sf s.c.}(T^{+}_{\sf melt})$, and the glass transition temperature $T_g$. Given the quality of the viscosity fit of Eq. (\ref{eq:visc1}) (Figure \ref{Collapse.}), there is little change in the estimated value of  $\Delta T_h$  when using either Eq. (\ref{scale-c-peak}) or Eq. (\ref{scale=1}).

There is an additional celebrated aspect of the heat capacity (alluded to in Section \ref{sec:config}) that we must touch on here for completeness. In our framework, as the system is progressively supercooled to lower temperatures, the support to the local reduced probability density from the higher temperature (high $T'$) states of the equilibrium system diminishes. Thus, the extrapolated entropy density of the supercooled fluid may approach that of the solid. Indeed, a well known extrapolation of the heat capacity data to temperatures below the experimentally measured $T_g$ has long suggested \cite{Walter} an impinging ``entropy crisis" mandates below the ``Kauzmann temperature'' 
$T_{K}$ at which the ascertained entropy of the glass becomes, paradoxically, lower than that of the equilibrium crystal.

\subsection{Structure} 
\label{sec:structure}

\begin{figure*}[t]
\includegraphics*[scale=0.38]{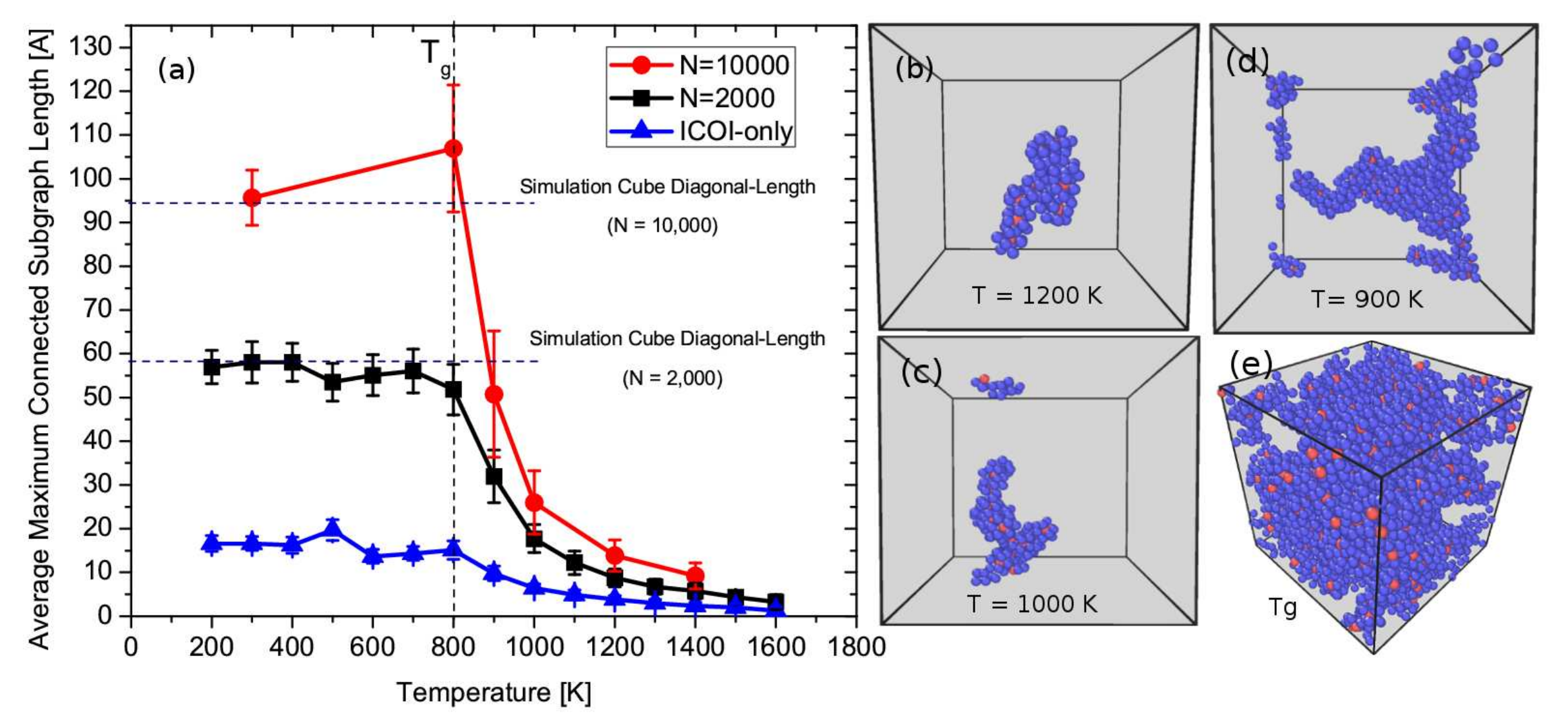}
\caption{(Color Online) Reproduced from \cite{Ryan1}. (a) Voronoi analysis (including that for varying system sizes) illustrating the increase of icosahedral backbone in simulated Cu$_{64}$Zr$_{36}$ subject to periodic boundary conditions. [(b) - (e)] Simulation snapshots at various temperatures. The percolating backbone starts to form at $T= T_{A} \sim 1600$ K.  The value of this temperature is consistently given by Eq. (\ref{TA:eqn}) \cite{critical} with the same value of $\overline{A}$ that fits the viscosity data (Eq. (\ref{eq:visc1}), Figure \ref{Collapse.}).}
\label{fig:ClusterLength}
\end{figure*}

\begin{figure}
\includegraphics[width=1 \columnwidth, height=.35 \textheight, keepaspectratio]{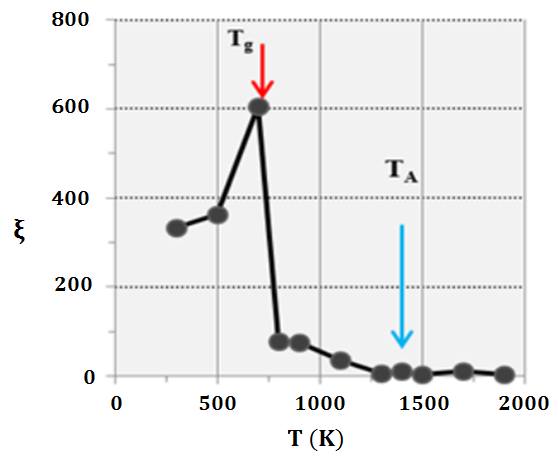}
\caption{(Color Online). Reproduced from \cite{shear1}.
Below $T_A$, metallic glasses may resist applied external shear \cite{shear1,shear2}. 
Shown is a plot of the shear penetration depth $\xi$ deduced from simulation of NiZr$_2$, of as a function of the temperature. Generally, the shear penetration depth starts to monotonically increasing from atomic scales once the temperature drops below a threshold temperature $T_A$. (At temperatures lower than the glass transition temperature $T_g$, the time required for the system to reach the rigid metastable glassy state exceeds the typical observation time.) The value of $T_A$ is consistent with that suggested by Eq. (\ref{TA:eqn}) with the value of $\overline{A}$ that fits the viscosity data (Eq. (\ref{eq:visc1}), Figure \ref{Collapse.}).}
\label{Lengthscale.}
\end{figure}

To examine the structure of supercooled liquids, we may invoke, once again, Eq. (\ref{Qavglong}) for chosen local structural quantities $Q$ (e.g., bond orientation order parameters \cite{BO} or local pair distribution functions). We underscore that apart from the conventional state variables, the ellipsis in Eq. (\ref{Qavglong}) may also include external stress, etc. 
Eq. (\ref{Qavglong}) will yield a relation between the average local structures within the supercooled liquid to those of the equilibrium crystal or fluid
(or their coexistence in the equilibrium ${\cal{PTEI}}$ region)
at different energy, particle densities, and pressures (as well as imposed local external stresses, etc.). As expected from Eq. (\ref{pp'}), in different regions of space, locally, the supercooled fluid may look like an equilibrium solid below its melting temperature or an equilibrium fluid of varying local energy densities, chemical potentials, pressures, etc. Eq. (\ref{pp'}) may, in this case, be interpreted as sampling different local equilibrium fluid and solid regions \cite{local_structure}. Experiments \cite{3D_Structure} indeed find a distribution of local crystal-like structures. The size of the local structures found \cite{3D_Structure} is compatible with the ${\sf n} \sim 15-135$ atom clusters suggested by the value of $\overline{A} \sim 0.1$ (e.g., Eq. (\ref{nd})). Consistent with experiment \cite{3D_Structure}, the absence of sharp Bragg peaks in such systems containing approximate local crystal-like structures should come as no surprise for these supercooled liquids that exhibit ``short range'' (i.e., typical nearest neighbor inter-atomic distance scale)  and ``medium range'' (several interatomic distances- characteristic nanometer scale) orders. Finite size broadening (scaling, in Fourier space, as the reciprocal of the linear size (i.e., as ${\cal{O}}({\sf n}^{-1/3})$)) of the structure factor peaks is extremely appreciable for such small (${\cal{O}}({\sf n})$ particle) clusters even if they were small finite size fragments of an ideal crystal. Importantly, as underscored in Section \ref{consequences}, {\it the very same distribution $P$ in Eq. (\ref{Qavglong}) that provided the thermodynamic and certain dynamic averages also yields all average structural (and any other) observables}.  

Using Eq. (\ref{Qavglong}), we may further also take $Q$ to be potential energy contributions or local stress components. Given, e.g., a local atomic configuration, we may evaluate each of the (short range) local contributions to its energy density. By virtue of being a sum of such local terms, the local energy density satisfies Eq. (\ref{Qavglong}). For instance,   short range potential energy contributions associated with the local atomic configurations must, self consistently, match  those given by the conditional probability $P$. Thus, e.g., Eq. (\ref{pp'}) implies that the distribution of local pair interaction energies is broadened by comparison to that in the equilibrium system by the conditional probabilities $P$. Such considerations will allow us to self consistently interpret general local mean potential energy densities, stresses, and the like as their respective weighted averages over a distribution of local equilibrium 
solid-like crystalline and equilibrium liquid type contributions.   

 Interestingly, the temperature ($T_A$) beyond which strong deviations from Arrhenius dynamics is consistent with the scaling relation
 \cite{critical} 
 \begin{eqnarray}
 \label{TA:eqn}
 T_A \sim \frac{T_{\sf melt}}{1- \overline{A}}. 
 \end{eqnarray} 
 The physical intuition underlying Eq. (\ref{TA:eqn}) is similar to the one that we employed for the specific heat  (Eq. (\ref{scale-c-peak})). That is,  given the distribution of Eq. (\ref{cluster}), as it being cooled down, the system will start ``sense'' the equilibrium solid-like states at a temperature $T_A$ where $T_{A} - \sigma_{T_A} (= T_{A}(1- \overline{A})) = T_{melt}$  yielding Eq. (\ref{TA:eqn}). Empirically, as its temperature drops below a crossover temperature $T_A$, the supercooled fluid starts to develop structural correlations. These correlations are seen by, e.g., an increasing icosahedral backbone in simulated metallic fluids (Figure \ref{fig:ClusterLength}) as found  by Voronoi analysis \cite{Ryan1,Ryan2} as well as an increased elastic rigidity evinced by the ability to resist shear over growing length scales (the supercooled fluid  ``shear penetration depth'' \cite{structure_ML,shear1,shear2}), see Figure \ref{Lengthscale.}, as the system is progressively supercooled to lower temperatures \cite{shear1,shear2}. 

 \subsection{The ``ideal glass''}
 We now close our circle of ideas and return to a concept that was shortly  mentioned in the Introduction. 
 Inspired by Kauzmann's suggestion of an ``entropy crisis'' \cite{Walter} (briefly discussed in Section \ref{sec:thermo}), Gibbs and DiMarzio \cite{GM} further postulated that a bona fide transition into (what has since been termed) an ``ideal glass'' may exist. Such a putative state was suggested to remain stationary forever and never crystallize. Discussions of this ideal state underlie very extensive work  \cite{Cavagna,BB}. In the context of the results that we derived this paper, so long as the conditions underlying our Theorem are satisfied, Eq. (\ref{pp}) must hold. Since this state is assumed to be of a divergent lifetime, the corrections discussed in Sections \ref{finite_time} and \ref{locality} may not appear. In particular, whenever the conditions for our Theorem are satisfied, if the ``ideal glass'' is not an equilibrium state governed by control parameters in the system Hamiltonian (e.g., the external magnetic field for a ferromagnet undergoing hysteresis as discussed in Section \ref{sect:proof_eqpp}) then the conditional probability $P$ will not be of a delta function form and, as seen from Eq. (\ref{non_equil_long}), {\em its correlations will not be clustered}. That is, such a putative state will exhibit system spanning connected correlations. If the conditions for our Theorem are not satisfied (as in e.g., many body localized \cite{MBL,MBL_rev} and scar \cite{scar,scar_rev} states, see also footnote [13]), a putative ``ideal glass'' state with clustered  correlations is possible.

 \section{Conclusions}
 A central result of the current work is encapsulated in our Theorem of Eq. (\ref{pp}) for the general form of any stationary reduced probability density. This theorem implies that {\it all} few body observables in a stationary system will be related to those in the equilibrium system by simple weighted averages (Eq. (\ref{Qavglong})) via the conditional probability distribution $P$ that depends only on state variables. We illustrated that, apart from special states (such as those in
 Sections \ref{decoupled_solvable} and \ref{1d-decoupled_particles}),
 the covariance of Eq. (\ref{non_equil_long}) will generally not vanish unless the system is in a standard thermal state. That is, generically, perfectly static states with clustered correlations are conventional equilibrium states. By our theorem, any deviations from the form of Eq. (\ref{pp}), including those necessary to allow for states other than those of usual equilibrium  to exhibit clustered correlations, will no longer be perfectly stationary. 

 Armed with the above formal result, we then applied Eq. (\ref{Qavglong}) to the  analysis of supercooled liquids. Indeed, although supercooled fluids ultimately crystallize, they may remain nearly stationary in that state for  exceedingly long times 
 (e.g., ${\cal{O}}(10^{10})-{\cal{O}}(10^{17})$ times the microscopic atomic time scales, see Section \ref{finite_time}). We discussed structural, thermodynamic, and dynamic consequences of our general form of the reduced probability densities and the expectation values that they imply. A simple picture embodying our approach is that of local regions of the supercooled fluid that look similar to those of the equilibrium system at different temperatures, chemical potentials, etc. The distribution of these equilibrium state variables plays a key role. Locally, some regions of the supercooled fluid may appear similar to small low temperature crystals (consistent with experiment \cite{3D_Structure}) while others emulate the higher temperature equilibrium fluid. Consequently, the excitations of the system and its dynamics may viewed as arising from  equilibrium solid and fluid-type contributions of varying state variables. A prediction of our work is that the single particle velocities in supercooled fluids may deviate (Eq. (\ref{no_Maxwell_dist})) from those of the Maxwell-Boltzmann form. The experimentally measured velocity distribution will then enable the extraction of the conditional probability $P$ (Eq. (\ref{Pfromv})). The degree of the latter deviation relates, in a precise way, to the viscosity of the system and all of its local few body observables (Eq. (\ref{viscosityfromv})). We further elaborated on a possible general qualitative relation that links the deviations between the cooling and heating traces of the specific heat and the viscosity (see Figure \ref{Specific_heat.}). The conditional probabilities $P$ appearing in Eqs. (\ref{pp}, \ref{pp'}) may allow for the prediction of numerous properties of supercooled fluids in a unified way. We outlined constraints implied by our theorem on the earlier conjectured  ``ideal glass'' state at low temperatures. As we explained in some depth, in our approach, the more elevated equilibrium melting transition temperature might, instead, be associated with the dynamics, thermodynamic, and structural properties of supercooled liquids. Such a suggestion differs from standard notions. Indeed, in mean-field type theories, supercooled liquids lie on a metastable branch of the free energy function which is featureless at the equilibrium melting temperature. In most theories of supercooled liquids \cite{Cavagna,BB,Steve_Gilles}, temperatures other than the equilibrium melting temperature are important. 
 
 Although Eq. (\ref{Qavglong}) was posited in earlier work \footnote{In Refs. \cite{1parameter,PS,longrange,critical,local_structure}, it was assumed that the {\it full many body} system remains in the supercooled state for divergent times. In actual experiments, the supercooled system exchanges heat with its surroundings and transitions into the crystalline equilibrium state at times $t_{\sf xtal}$ set by the driving free energy for nucleation and growth \cite{KG}. If heat could not be exchanged with the environment then any nucleation might heat up the system- thus in turn slowing crystallization and possibly remelting the sample. In such an idealized setting, the supercooled system might avoid crystallization for a very long time (potentially even at divergent times as implicitly assumed in \cite{1parameter,PS,longrange,critical,local_structure} that considered ideal perfectly isolated metastable supercooled liquids to remain unchanged). In the current work, we focused, by contrast, on the empirically pertinent near-stationarity of few body observables. Local uncorrelated noise from the environment may (and does) trigger nucleation and bona-fide equilibriation to the crystalline state (at sufficiently long times) and can thus inhibit long-range correlations.}, our investigation goes far beyond that of Refs. \cite{1parameter,PS,longrange,critical,local_structure}. 
We do not assume that long-range correlations exist in the supercooled system. Indeed, locality and (near) stationarity of few body observables form the pillars of our analysis.

 {\bf Acknowledgments.} I am very grateful to Shivaji L. Sondhi, John T. Chalker, Fangzheng Chen, Kenneth F. Kelton, Louk Rademaker, Ylias Sadki, Giorgi Tsereteli, Jing Xue,  Takato Yoshimura, and other members of the Oxford glass group for valuable discussions and much ongoing work. I also wish to thank John T. Chalker, Shmuel Nussinov, and Shivaji L. Sondhi for a careful reading of the manuscript. My work at Oxford, where this research was carried out, was supported by a Leverhulme Trust International Professorship grant [number LIP-2020-014].

\appendix

\section{Clustered correlations in a non-stationary system}
\label{example=cluster}
To illustrate how shorter correlation lengths may appear in tandem with more rapid dynamics, we consider (for a system at an inverse temperature $\beta$) the set of clustered two-body ($n=2$) conditional probability densities given by
\begin{eqnarray}
\label{r2w}
\rho_2(i,j) =  {\cal{N}}(\beta) \int d\beta'   \Big(w_{ij}(\beta') R_2(\beta|\beta')  \rho_{{\sf eq.},2}(\beta') \nonumber
\\ + (1- w_{ij}(\beta'))  \int d \beta'' R_1(\beta'| \beta'') \rho_{{\sf eq.}, 1}(i)(\beta'') \nonumber
\\ \times \int d \beta''  R_1 (\beta'|\beta'') \rho_{{\sf eq.}, 1}(j)
(\beta'') \Big),
\end{eqnarray}
with $R_{1}$ and $R_{2}$ arbitrary conditional probability distributions, and ${\cal{N}}$ a normalization factor that depends only on the state variables of the system (i.e., does not involve any of its phase space coordinates or other observables). For simplicity, we discuss  what occurs when $w_{ij}$ is only a function of the spatial coordinates of particles $i$ and $j$ (and not their momenta). To ensure clustering, in Eq. (\ref{r2w}), the function $w_{ij}$ must tend to zero when the inter-particle distance
$r_{ij} \equiv |{\bf x}_i - {\bf x}_j|  \to \infty$. The covariance (similar to that in Eq. (\ref{non_equil_long})) between asymptotically distant local operators $A$ and $B$ will then  
vanish. 
Eq. (\ref{r2w}) can be generalized to reduced $n>2$ particle probability densities. 
In Eq. (\ref{r2w}), any function $w_{ij}$ depending on the coordinates of moving particles (by contrast to that whose arguments are the stationary positions of spins in a lattice) 
will, generally, be a function of time. All other things kept constant, the more rapid the decay of $w_{ij}$ with increasing inter-particle distance $r_{ij}$, 
the larger time derivative of $w_{ij}$ will be. As a concrete example, we consider
\begin{eqnarray}
\label{w-decay}
  w_{ij} \sim e^{-r_{ij}/\xi},  
\end{eqnarray}
to be of the qualitative form of a typical connected correlation function 
\footnote{The astute reader will indeed  recognize that given Eq. (\ref{r2w}), this form of $w_{ij}$ implies that the connected two-body correlations indeed decay with the distance with a correlation length $\xi$.} and demonstrate that a small correlation length $\xi$ in Eq. (\ref{w-decay}) may enable large time derivatives of $w_{ij}$ and thus to sizable temporal changes of the reduced two-particle probability density of Eq. (\ref{r2w}). Physically, a reduced  probability density of the form of Eqs. (\ref{r2w}, \ref{w-decay}) for which the covariance between local  observables decays with their distance can arise from, e.g., local noise (either external or self generated) that is spatially uncorrelated across the system in a metastable state. To ensure clustering, a spatial dependence such as that in Eqs. (\ref{r2w},\ref{w-decay}) must, generally, be present. This is an example of the earlier noted broader maxim: the more rapid the spatial fluctuations of the reduced probability density are about those of Eq. (\ref{pp}), the quicker its time fluctuations may be. Beyond the general results outlined in Section \ref{locality}, further approximate  considerations suggest how such contributions to the spatial fluctuations might indeed go hand in hand with those leading to larger temporal fluctuations and finite lifetime of the (nearly stationary not yet permanent) metastable state in simple example systems. 
Henceforth, we will largely employ a classical language. Nonetheless, all of our considerations apply equally well for quantum systems. Specifically, 
With \(m_i\) denoting the mass of the \(i\)-th particle, $\bf{x}_{i}$ and ${\bf p}_{i}$ its position and momentum, and $V_{ij} \equiv V(\mathbf{x}_i - \mathbf{x}_{j})$ the pairwise interaction potential between the \(i\)-th and \(j\)-th particles, the BBGKY hierarchy \cite{BBGKY1,BBGKY2,BBGKY3,BBGKY4,BBGKY5} for the reduced probability density   
explicitly reads 
\begin{eqnarray}
\label{BBGKY-eq}
 \Big(\frac{\partial}{\partial t} + \sum_{i=1}^{n} \frac{{\bf p}_i}{m_i} \cdot {\bf \nabla}_i -  \sum_{i=1}^{n} \sum_{j\neq i} {\bf \nabla}_i V_{ij}  \cdot \frac{\partial }{\partial {\bf p}_i} \Big)  \rho_{n}  = \nonumber
\\  \sum_{i=1}^{n} \int d^{d}{\bf x}_{n+1} \int d^{d} {\bf p}_{n+1} \left( {\bf \nabla}_i V_{i,n+1} \right) \cdot \frac{\partial \rho_{n+1}}{\partial {\bf p}_i}. 
\end{eqnarray}
We briefly remind the reader of the  simple origin of this hierarchy so that the physical content of its application to our problem and of the approximations that we will shortly make will become clear: Interactions with an external ($(n+1)$-th) particle alter the momenta of the $n$ particles (for which the reduced probability density $\rho_n$ is defined) resulting in a contribution (appearing on  the righthand side of Eq. (\ref{BBGKY-eq})) to the total time derivative of $\rho_n$ as computed with the chain rule applied to the phase space coordinates (lefthand side of Eq. (\ref{BBGKY-eq})). In the large $\xi$ limit of Eq. (\ref{w-decay}) (i.e., whenever $w_{ij}$ is a constant), the reduced probability density of Eq. (\ref{r2w}) will become stationary (since it will be a function of only (time independent) equilibrium reduced probability densities).  
In the latter $\xi \to \infty$ limit, we will regain Eq. (\ref{pp}) for general (i.e., not only local) reduced probability densities. It is only when $w_{ij}$ is not a constant that the time derivative of Eq. (\ref{r2w}) can be finite. Given Eqs. (\ref{r2w}, \ref{w-decay}), the non-vanishing contributions to $\partial_t  \rho_{n=2}(i,j)$ from its convective phase space derivative in Eq. (\ref{BBGKY-eq}),
$\Big(-\sum_{i=1}^{n=2} \frac{{\bf p}_i}{m_i} \cdot \nabla_i +  \sum_{i=1}^{n=2} \sum_{j\neq i} {\bf \nabla}_i V_{ij}  \cdot \frac{\partial }{\partial {\bf p}_i} \Big)  \rho_{n=2}(i,j)$, 
involve the derivative of $w_{ij}$ with respect to relative distance $r_{ij}$ between particles $i$ and $j$ times $dr_{ij}/dt$ (i.e., times the relative velocity between the two particles). Thus, excusing the external interaction term on the righthand side of Eq. (\ref{BBGKY-eq}), since $dw_{ij}/dr_{ij} \sim - w_{ij}/\xi$, the partial time derivative of 
$\rho_2(i,j)$ will similarly scale monotonically with $\xi$. Hence, more rapid spatial fluctuations of the probability density progressively may naturally go hand in hand with faster dynamics. The temporal derivative of the reduced probability density will be amended by the above omitted external interaction term appearing on the righthand side Eq. (\ref{BBGKY-eq}).  Almost identical analysis may be pursued for reduced probability densities that constitute extensions of Eq. (\ref{r2w}) to larger $n>2$ values. Similar conclusions are also suggested by applying the chain rule to compute the time derivatives of reduced probability densities other than those of Eqs. (\ref{r2w}, \ref{w-decay}), including those of higher particle number, that fluctuate about the purely static Eq. (\ref{pp}). As discussed in Section \ref
{locality}, beyond the approximate considerations that we invoked in this Appendix for rather specific example systems, far more generally, in the limit of asymptotically long-lived  metastable states, the correlation length $\xi$ of Eq. (\ref{w-decay}) must be monotonic in the  lifetime of such states. 

\section{Clustered states governed by local nearly commuting Hamiltonians}
\label{nearly-commute}
In Section \ref{locality} and Appendix \ref{example=cluster}, we discussed clustered correlations. Here, we will briefly expand on quantum systems. The example of Eq. (\ref{trivexample}) is, of course, rather special since there are $N$ local symmetries (i.e., the individual $\{H_{i}\}_{i=1}^{N}$) that each commute with the full system Hamiltonian $H$ and may be simultaneously diagonalized. As a slightly more realistic situation, similar to Section \ref{locality}, we consider a system displaying clustered, local correlations
(so that the probability densities over distant regions are uncorrelated and factorize) with the Hamiltonian a sum of local non-commuting terms $\{ H_{i} \}_{i=1}^{N}$ with bounded decaying commutators (as is the case when the Lieb-Robinson bounds apply \cite{LR1,LR}), 
\begin{eqnarray}
\label{LReqn}
  ||[H_{i}, H_{j} ] || \le C e^{-|i-j|/\ell},
\end{eqnarray}
with constant $C$ and $\ell$. 
We next consider a local probability density having its support in the region ${\cal{R}}_{i}$ where $H_{i}$ operates which is of the form  $\rho_{1,i} = f(H_{i})$
with $f$ a general function. In such a case, as observed from 
Eq. (\ref{LReqn}), the commutator $[H,\rho_{1,i}]$ will become bounded from above by a function decaying exponentially in the linear size of the region ${\cal{R}}_{i}$. Up to these exponential corrections, the reduced probability density $\rho_{1,i}$ must then be of the form of Eqs. (\ref{pp}, \ref{pp'}). 

\section{Clustered static  approximations to the probability density}
\label{ap:approx}

        If Eq. (\ref{pp}) holds for a local reduced probability density of any particular particle number $n$ then (as is evident by performing further partial traces) it clearly further holds for all local $n_{a}<n$ reduced probability densities. Let us next consider general systems and ask what occurs when a Kirkwood type \cite{kirkwood1,kirkwood2,Singera} closure 
        relation expresses the reduced probability densities for more than $n$ particles in terms of those of  $n_{a} \le n$ particles and when the correlations are local \footnote{In the simplest (Kirkwood) approximation, the associated
        correlation function of three particles is equal to the product of the three pair correlation functions, $g({\bf{x}}_1,{\bf{x}}_2,{\bf{x}}_3) = g({\bf{x}}_1,{\bf{x}}_2)g({\bf{x}}_2,{\bf{x}}_3)g({\bf{x}}_1,{\bf{x}}_3)$.  Such an approximation neglects subtle higher body correlations.}. That is, we wish to examine a clustering  expansion truncated at finite ($n-$th) order so that there are no connected clusters involving more than $n$ particles. The system probability density may then be  approximated by
        \begin{eqnarray}
        \label{clustering_rel}
         \rho_{\Lambda} 
         \sim {\cal{F}}(\{\rho_{n_a}\}),
        \end{eqnarray}
        where ${\cal{F}}$ is a function that may assume different forms depending on the clustering approximations used \cite{Singera}. We further impose a constraint for the reduced probability densities of $n_a \le n$ particles wherein whenever if any of the particles are far apart, the probability densities factorize 
        (within an error that tends to zero for large  interparticle separations).
        In what follows, we will 
        regard this as a given candidate state. 
        In 
        Eq. (\ref{clustering_rel}), ${\cal{F}}$ is a 
        function of {\it local} reduced probability densities having $n_a \le n$ particles. When both Eq. (\ref{pp}) and Eq. (\ref{clustering_rel}) hold, the full many body probability density can be written as
        \begin{eqnarray}
\label{pp_prod}
 \rho_{\Lambda} \sim 
&& {\cal{F}}\Big( \Big\{ \int d\beta'_a ~d\mu'_a \ldots ~~P(\beta, \mu, \ldots| \beta_a', \mu_a', \ldots ) \nonumber
\\  && \times ~~\rho_{{\sf eq.},n_a}(\beta_a',\mu_a', \ldots) \Big\} \Big).
\end{eqnarray}
Formally, as we earlier demonstrated, when ensemble equivalence applies, all exactly static many body probability densities must be of the form of Eq. (\ref{longpp+}). Eq. (\ref{pp_prod}) is a particular subset of static states that may approximate $\rho_{\Lambda}$ when Eq. (\ref{clustering_rel}) applies.
The example of Section \ref{decoupled_solvable} constitutes a rather exceptional case in which clustering is simply satisfied for general $P$. In the spin model of Section \ref{decoupled_solvable}, the many body probability density $\rho_{\Lambda}$ is given by such a state with ${\cal{F}}$ being the single fully symmetric monomial- the product of the $N$ single particle ($n=n_a=1$) reduced probability densities. 
Generally, by virtue of the approximate decomposition of the many body probability density of Eq. (\ref{pp_prod}) into local finite $n_{a} \le n-$body probability densities, it follows that if the equilibrium system has local connected correlations then the standard deviation of, e.g., the global system energy density (i.e., the standard deviation of the total system energy divided by the particle number) will vanish in the thermodynamic limit as
\begin{eqnarray}
\label{scalesq}
\frac{\sigma_{E}}{N} = {\cal O}(N^{-1/2}).
\end{eqnarray}
The general scaling relation of Eq. (\ref{scalesq}) immediately follows from a direct calculation that is, once again, nearly identical to that in Section \ref{decoupled_solvable}. If the full system Hamiltonian can be expressed as a sum of {\it local few body} Hamiltonians,
\begin{eqnarray}
H = \sum_{i} H_{i},
\end{eqnarray}
then the variance of the energy density ($E/N$) is given by
\begin{eqnarray}
\label{sE}
\frac{\sigma_{E}^2}{N^2} = \frac{1}{N^2} \sum_{ij} G_{ij}, 
\end{eqnarray}
where $G_{ij}$ is the covariance between the local terms in the Hamiltonian, 
\begin{eqnarray}
\label{sE1}
G_{ij}=  \langle H_{i} H_{j} \rangle - \langle H_i \rangle \langle H_j \rangle.
\end{eqnarray}
We reiterate that the local operators $H_{i}$ and $H_{j}$ have their support on a finite number of particles. If all few body reduced probability densities $\rho_{n}$ are {\it local} (i.e., do not have non-vanishing connected correlations for far separated particles) then it follows that for each $i$, the covariance $G_{ij}$ is non-vanishing only for a finite number of values of $j$. This, in turn, implies that the variance of Eq. (\ref{sE}) scales as $1/N$ (as in Eq. (\ref{scalesq})).  
This scaling is a consequence of the locality of the reduced few body equilibrium probability densities in Eq. (\ref{pp}) and does not hinge on the validity of the 
closure approximations of Eq. (\ref{clustering_rel}). In a similar vein, whenever the system exhibits a clustering of correlations (so that $G_{ij}$ decays sufficiently rapidly that the sum in Eq. (\ref{sE}) tends to zero in the thermodynamic limit), whether the system is in a static state or a time-varying one, the global energy density will exhibit a sharp delta function like distribution in its energy density (with analogous relations for the distributions of other intensive state variables). 

Even when relations of the type of Eqs. (\ref{clustering_rel}, \ref{pp_prod}) 
 for the probability density of the full many body system $\Lambda$ do not hold, whenever Eq. (\ref{pp}) applies, the local reduced $n-$particle probability density will be time independent. Consequently, when Eq. (\ref{pp})  is satisfied,  the system will appear to be stationary for conventional few particle (i.e., finite order $\le n$ body) correlations. 
    While the functions appearing on the righthand side of Eq. (\ref{clustering_rel}) and in Eq. (\ref{pp_prod}) are exactly  stationary, they generally do not exactly describe the full probability density of metastable states that are of a finite lifetime. As we next explain, deviations from the exactly stationary Eqs. (\ref{pp}, \ref{pp_prod}) may indeed call for the appearance of additional corrections to a non-equilibrium $\rho_{\Lambda}$ that may be non-stationary. 

Any dynamics of the full system (i.e., a non-vanishing $\partial \rho_{\Lambda}/\partial t$) may only arise from a violation of Eq. (\ref{pp}) and/or of the assumption of a factorization closure relation of Eq. (\ref{clustering_rel}) for arbitrary configurations. 
As illustrated in the example of Section \ref{decoupled_solvable}, for the simple case of decoupled local interactions with locally separable probability densities, Eqs. (\ref{pp},\ref{clustering_rel}) lead to an exact static many body probability density. In the latter example, any reduced $n$-particle probability density is just a product of reduced single particle probability densities. In more  general systems, when no such decoupling occurs, non-trivial connected correlations will lead to corrections to the clustering expansion of Eq. (\ref{clustering_rel}). These corrections may decay exponentially with the inter-particle distances. Given Eq. (\ref{eq:non-local}), complete clustering can only arise if, in Eq. (\ref{pp}), $P$ is a delta function. As Eq. (\ref{eq:non-local}) illustrates, any other nontrivial conditional probability $P$ having a finite standard deviation cannot, typically, describe a completely stationary state if the latter exhibits a clustering of its correlations.

    \bibliographystyle{natbib}
\bibliographystyle{ieeetr}
\bibliography{ref.bib}

\begin{thebibliography}{100}

\bibitem{Cavagna}
A.~Cavagna, ``Supercooled liquids for pedestrians,'' {\em Physics Reports}, vol.~476, p.~51, 2009.

\bibitem{BB}
L.~Berthier and G.~Biroli, ``Theoretical perspective on the glass transition and amorphous materials,'' {\em Reviews of Modern Physics}, vol.~83, p.~587, 2011.

\bibitem{precambrian}
A.~D. Fowler, M.~Berger, M.~Shore, M.~I. Shore, and J.~R. Jones, ``Supercooled rocks: development and significance of varioles, spherulites, dendrites and spinifex in archaean volcanic rocks, abitibi greenstone belt, canada,'' {\em Precambrian Research}, vol.~115, p.~311, 2002.

\bibitem{Angell}
C.~A. Angell, ``Formation of glasses from liquids and biopolymers,'' {\em Science}, vol.~267, p.~1924, 1995.

\bibitem{RFOT1}
T.~R. Kirkpartick and D.~Thirumalai, ``p-spin-interaction spin-glass models: Connections with the structural glass problem,'' {\em Phys. Rev. B}, vol.~36, p.~5388, 1987.

\bibitem{RFOT2}
T.~R. Kirkpatrick and P.~G. Wolynes, ``Connections between some kinetic and equilibrium theories of the glass transition,'' {\em Phys. Rev. A}, vol.~35, 1987.

\bibitem{RFOT3}
T.~R. Kirkpatrick, D.~Thirumalai, and P.~G. Wolynes, ``Scaling concepts for the dynamics of viscous liquids near an ideal glassy state,'' {\em Phys. Rev. A}, vol.~40, p.~1045, 1989.

\bibitem{RFOT4}
T.~R. Kirkpatrick and D.~Thirumalai, ``Random solutions from a regular density functional hamiltonian: a static and dynamical theory for the structural glass transition,'' {\em J. Phys. A: Math. Gen.}, vol.~22, p.~L149, 1989.

\bibitem{RFOT_review}
G.~Biroli and J.-P. Bouchaud, ``The rfot theory of glasses: Recent progress and open issues,'' {\em arXiv}, 2022.

\bibitem{Annrev17}
P.~Charbonneau, J.~Kurchan, G.~Parisi, P.~Urbani, and F.~Zamponi, ``Glass and jamming transitions: From exact results to finite-dimensional descriptions,'' {\em Annual Review of Condensed Matter Physics}, vol.~8, p.~265, 2017.

\bibitem{Parisi_book}
G.~Parisi, P.~Urbani, and F.~Zamponi, {\em Theory of Simple Glasses}.
\newblock Cambridge University Press, 2020.

\bibitem{david}
D.~R. Reichman and P.~Charbonneau, ``Mode-coupling theory,'' {\em Journal of Statistical Mechanics}, vol.~2005, p.~P05013, 2005.

\bibitem{mode_coupling}
M.~C. Janssen, ``Mode-coupling theory of the glass transition: A primer,'' {\em Frontiers in Physics}, vol.~6, p.~97, 2018.

\bibitem{dyn_fac}
J.~P. Garrahan and D.~Chandler, ``Coarse-grained microscopic model of glass formers,'' {\em Proceedings of the National Academy of Science}, vol.~100, p.~9710, 2003.

\bibitem{dk}
D.~Kivelson, S.~A. Kivelson, X.~Zhao, Z.~Nussinov, and G.~Tarjus, ``A thermodynamic theory of supercooled liquids,'' {\em Physica A}, vol.~219, p.~27, 1995.

\bibitem{tarjus}
G.~Tarjus, S.~A. Kivelson, Z.~Nussinov, and P.~Viot, ``The frustration based approach to supercooled liquids and the glass transition: a review and critical assessment,'' {\em Journal of Physics: Condensed Matter}, vol.~17, p.~R1143, 2005.

\bibitem{nab}
Z.~Nussinov, ``Avoided phase transitions and glassy dynamics in geometrically frustrated systems and non-abelian theories,'' {\em Physical Review B}, vol.~69, p.~014208, 2004.

\bibitem{dyre}
J.~C. Dyre, ``The glass transition and elastic models of glass-forming liquids,'' {\em Rev. Mod. Phys.}, vol.~78, p.~953, 2006.

\bibitem{Steve_Gilles}
S.~A. Kivelson and G.~Tarjus, ``In search of a theory of supercooled liquids,'' {\em Nature Materials}, vol.~7, p.~831, 2008.

\bibitem{Walter}
W.~Kauzmann, ``The nature of the glassy state and the behavior of liquids at low temperatures,'' {\em Chemical Reviews}, vol.~43, p.~219, 1948.

\bibitem{GM}
J.~H. Gibbs and E.~A. DiMarzio, ``Nature of the glass transition and the glassy state,'' {\em J. Chem. Phys.}, vol.~28, p.~373, 1958.

\bibitem{1parameter}
Z.~Nussinov, ``A one parameter fit for glassy dynamics as a quantum corollary of the liquid to solid transition,'' {\em Philosophical Magazine}, vol.~97, p.~1509, 2017.

\bibitem{PS}
N.~B. Weingartner, C.~Pueblo, F.~Nogueira, K.~F. Kelton, and Z.~Nussinov, ``A phase space approach to supercooled liquids and a universal collapse of their viscosity,'' {\em Frontiers in Materials}, vol.~3, p.~50, 2016.

\bibitem{longrange}
Z.~Nussinov, ``Macrosopic length correlations in non-equilibrium quantum systems and their possible experimental realizations,'' {\em Nuclear Physics B}, vol.~953, p.~114958, 2020.

\bibitem{critical}
N.~B. Weingartner, C.~Pubelo, K.~F. Kelton, and Z.~Nussinov, ``Critical assessment of the equilibrium melting-based, energy distribution theory of supercooled liquids and application to jammed systems,'' p.~arXiv:1512.04565, 2015.

\bibitem{local_structure}
Z.~Nussinov, N.~B. Weingartner, and F.~S. Noguiera, {\em Chapter titled ``The `glass transition' as a topological defect driven transition in a distribution of crystals and a prediction of a universal viscosity collapse'' in the book on ``Topological Phase Transitions and New Developments'' edited by Lars Brink, Mike Gunn, Jorge V. Jose, John Michael Kosterlitz, and Kok Khoo Phua}.
\newblock World Scientific, 2018.

\bibitem{Note1}
This includes, e.g., for the case of the supercooled liquid, any numerically obtained/experimentally measured long time average of the reduced probability density in the time interval $t_{\protect \sf xtal} \ge t \ge t_{\min }$.

\bibitem{Note2}
For the example of the supercooled liquid, this refers to the long time average of $\rho _{\Lambda }$ within the time interval $t_{\protect \sf xtal} \ge t \ge t_{\min }$ where the average few body observables are nearly static and the system appears to reach an effective equilibrium.

\bibitem{Note3}
Since at any time $t'$, the full many body probability density $\rho _{\Lambda }(t')$ is normalized, it satisfies Fubini's theorem. In other words, the order of the partial trace in Eq. (\ref {full}) and the integration over the time variable such as that associated with the averaged long time integral of Eq. (\ref {rlta}) can be interchanged. This interchange leads to Eq. (\ref {ft}) with the definition of Eq. (\ref {llta}) (and analogously leads to Eq. (\ref {trivw}) when noting that the averaged time integral of the full trace of $|\rho _{\Lambda }(t') e^{i \omega t'}|$ is also trivially normalized- thus similarly satisfying the condition for Fubini's theorem).

\bibitem{Note4}
As is well known, e.g. \cite {LL1}, for a system with $f$ degrees of freedom, there may be $(2f-1)$ conserved quantities when integrating the corresponding equations of motion (with a second order differential equation (and thus, apart from a trivial time shift, two constants of motion) for each degrees of freedom); our focus is on the symmetry borne additive conserved quantities that lead to intensive state variables in the thermodynamic limit.

\bibitem{Note5}
If the function ${\protect \cal P}$ is not smooth then Eq. (\ref {longpp+}) will be valid only if an extreme form of general ensemble equivalence (a condition for our Theorem)- the Eigenstate Thermalization Hypothesis (ETH) \cite {ETH1,ETH2}- holds. The ETH asserts that one can take the microcanonical ensemble in the limiting form of a single energy eigenstate. Whenever the ETH holds, the projection operator to each individual eigenstate inasmuch as local observables are concerned may be replaced by an equilibrium probability density set by the state variables $E', \{W'_{\alpha }\}$. In systems in which the ETH is not satisfied (e.g., those with many body localization \cite {MBL,MBL_rev} or in scar \cite {scar,scar_rev} states), Eq. (\ref {longpp+}) need not be satisfied.

\bibitem{Note6}
Observe that Eq. (\ref {cond_prob}) makes no assumptions about the state of the stationary system. The function ${\protect \cal P}(E',N', \{W'_{\alpha }\})$ may be a rather rapidly varying one. All that is important is that the equilibrium energy $E'$ is associated with a unique inverse temperature $\beta '$ for which $E' = U(\beta ')$ with $U$ the internal energy of the equilibrium system and that, similarly, the particle number $N'$ is uniquely determined by the chemical potential $\mu '$. In Eq. (\ref {pp'}), we will discuss situations (e.g., those of equilibrium phase coexistence) in which an extensive range of equilibrium energies $E'$ (finite range of equilibrium energy densities) is associated with a single inverse temperature $\beta '$.

\bibitem{Note7}
For the supercooled liquid, the non-stationarity of the full many body system is, e.g., evinced by the time dependent character of the spatially heterogeneous dynamics of the liquid \cite {Cavagna,BB,DH2,DH3,rotation}).

\bibitem{ETH1}
J.~M. Deutsch, ``Quantum statistical mechanics in a closed system,'' {\em Phys. Rev. A}, vol.~43, p.~2046, 1991.

\bibitem{ETH2}
M.~Srednicki, ``Chaos and quantum thermalization,'' {\em Phys. Rev. E}, vol.~50, p.~888, 1994.

\bibitem{Kelton_review}
K.~F. Kelton, {\em ``Theory of Nucleation and Glass Formation'' in book titled ``Metallurgy in Space, Recent results from ISS'' Edited by H-J. Fecht and M. Mohr, ``The Minerals, Metals, \& Materials Series''}.
\newblock Springer, 2022.

\bibitem{Ganorkar}
S.~Ganorkar, S.~Lee, Y.~H. Lee, T.~Ishikawa, and G.~W. Lee, ``Origin of glass forming ability of cu-zr alloys: A link between compositional variation and stability of liquid and glass,'' {\em Phys. Rev. Mater.}, vol.~2, p.~115606, 2018).

\bibitem{glycerol-TTT}
Q.~Wang, X.~Huang, W.~Guo, and Z.~Cao, ``Synergy of orientational relaxation between bound water and confined water in ice cold-crystallization,'' {\em Physical Chemistry Chemical Physics}, vol.~21, p.~10293, 2019.

\bibitem{KG}
K.~F. Kelton and A.~L. Greer, {\em Nucleation in Condensed Matter: Applications in Materials and Biology}.
\newblock Elsevier, 2010.

\bibitem{Note8}
As they must, the large $\tau $ averages of the reduced probability densities $\rho _{n,l.t.a.}$ and hence the conditional probabilities $P_{\protect \sf t}$ are nearly constant in ${\protect \sf t}$ whenever all few body observables exhibit no appreciable dynamics as is the case for supercooled liquids in their long lived metastable state.

\bibitem{Note9}
This explicitly follows from (i) the positivity of the correlation length $\xi $ and lifetime $\tau _{\ell }$, (ii) the divergence of the correlation length $\xi $ implied by Eq. (\ref {eq:non-local}) when the lifetime of the non-equilibrium system $\tau _{\ell } \to \infty $, and the above noted (iii) assumed well-defined function $\xi ^{-1}(\tau ^{-1}_{\ell })$ for asymptotically large $\tau _{\ell }$. Property (ii) implies that $\lim _{\tau _{\ell } \to \infty } \xi ^{-1}(\tau _{\ell }^{-1}) =0$. Property (i) then mandates that in the vicinity of the latter asymptotic $\xi ^{-1}=\tau _{\ell }^{-1}=0$ point, the function $\xi ^{-1}(\tau _{\ell }^{-1})$ must be monotonically non-decreasing.

\bibitem{superheat}
M.~Gallo, F.~Magaletti, and C.~M. Casciola, ``Thermally activated vapor bubble nucleation: The landau-lifshitz-van der waals approach,'' {\em Phys. Rev. Fluids}, vol.~3, p.~053604, 2018.

\bibitem{localeTplasma}
B.~Nold, T.~T. Ribeiro, M.~Ramisch, Z.~Huang, H.~W. Muller, B.~D. Scott, U.~Stroth, and the ASDEX Upgrade~Team, ``Influence of temperature fluctuations on plasma turbulence investigations with langmuir probes,'' {\em New Journal of Physics}, vol.~14, p.~063022, 2012.

\bibitem{Doyon}
B.~Doyon, ``Notes on generalized hydrodynamics,'' {\em SciPost Phys. Lect. Notes}, p.~18, 2020.

\bibitem{SG1}
M.~Mezard, G.~Parisi, and M.~A. Virasoro, {\em Spin glass theory and beyond}.
\newblock Singapore: World Scientific, 1987.

\bibitem{DH1}
H.~Sillescu, ``Heterogeneity at the glass transition: a review,'' {\em J. of Non-Crystalline Solids}, vol.~243, p.~81, 1999.

\bibitem{DH2}
M.~D. Ediger, ``Spatially heterogeneous dynamics in supercooled liquids,'' {\em Annual Review of Physical Chemistry}, vol.~51, p.~99, 2000.

\bibitem{DH3}
R.~Richert, ``Heterogeneous dynamics in liquids: fluctuations in space and time,'' {\em J. of Physics: Condensed Matter}, vol.~14R, p.~703, 2002.

\bibitem{rotation}
L.~J. Kaufmann, ``Heterogeneity in single-molecule observables in the study of supercooled liquids,'' {\em Ann. Rev. Phys. Chem.}, vol.~64, p.~177, 2013.

\bibitem{Note10}
In this application (and others like it), the lower $t'=0$ limit of the integrals corresponds, as in Section \ref {finite_time}, to the minimal waiting time $t=t_{\min }$ for the system to reach its long-lived nearly stationary state.

\bibitem{Alecu}
T.~I. Alecu, S.~Voloshynovskiy, and T.~Pun, ``The gaussian transform,'' {\em EUSIPCO2005, 13th European Signal Processing Conference}, 2005.

\bibitem{Nose}
S.~Nose, ``A unified formulation of the constant temperature molecular dynamics methods,'' {\em Journal of Chemical Physics}, vol.~81, p.~511, 1984.

\bibitem{Hoover}
W.~G. Hoover, ``Canonical dynamics: Equilibrium phase-space distributions,'' {\em Phys. Rev. A}, vol.~31, p.~1695, 1985.

\bibitem{Andersen_thermostat}
H.~C. Andersen, ``Molecular dynamics simulations at constant pressure and/or temperature,'' {\em The Journal of Chemical Physics}, vol.~72, p.~2384, 1980.

\bibitem{Langevin1}
W.~G. Hoover, A.~J. Ladd, and B.~Moran, ``High-strain-rate plastic flow studied via nonequilibrium molecular dynamics,'' {\em Phys. Rev. Lett.}, vol.~48, p.~1818, 1982.

\bibitem{Langevin2}
D.~J. Evans, ``Computer ‘‘experiment’’ for nonlinear thermodynamics of couette flow,'' {\em J. Chem. Phys.}, vol.~78, p.~3297, 1983.

\bibitem{Giorgi}
G.~Tsereteli~et al. {\em In preparation}.

\bibitem{non-Maxwell}
X.~Su, A.~Fischer, and F.~Cichos, ``Towards measuring the maxwell–boltzmann distribution of a single heated particle,'' {\em Front. Phys.}, vol.~9, p.~669459, 2021.

\bibitem{Goldstein}
M.~Goldstein, ``Viscous liquids and the glass transition: a potential energy barrier picture,'' {\em J. Chem. Phys.}, vol.~51, p.~3728, 1969.

\bibitem{Ryan2}
R.~Soklaski, V.~Tran, Z.~Nussinov, K.~F. Kelton, and L.~Yang, ``A locally-preferred structure characterizes all dynamical regimes of a supercooled liquid,'' {\em Philosophical Magazine}, vol.~96, p.~1212, 2016.

\bibitem{Kostya-elastic}
K.~Trachenko, ``Heat capacity of liquids: An approach from the solid phase,'' {\em Phys. Rev. B}, vol.~78, p.~104201, 2008.

\bibitem{3D_Structure}
Y.~Yang, J.~Zhou, F.~Zhu, Y.~Yuan, D.~J. Chang, D.~S. Kim, A.~Rana, X.~Tian, Y.~Yao, S.~J. Osher, A.~K. Schmid, L.~Hu, P.~Ercius, and J.~Miao, ``Determining the three-dimensional atomic structure of an amorphous solid,'' {\em Nature}, vol.~592, p.~7852, 2021.

\bibitem{volynes}
P.~Zhang, J.~J. Maldonis, Z.~Liu, J.~Schroers, and P.~M. Volynes, ``Spatially heterogeneous dynamics in a metallic glass forming liquid imaged by electron correlation microscopy,'' {\em Nature Communications}, vol.~9, p.~1129, 2018.

\bibitem{PVAC}
J.-L. Garden and H.~Guillou, {\em Thermodynamics of Glasses, in Encyclopedia of Glass Science, Technology, History, and Culture}.
\newblock Wiley, 2020.

\bibitem{Tang}
M.~B. Tang, W.~H. Wang, and J.~T. Zhao, ``Constant-volume heat capacity at glass transition,'' {\em Journal of Alloys and Compounds}, vol.~577, p.~299, 2013.

\bibitem{Note11}
Crystalline defects (e.g., dislocations) may drift following the application of a finite external stress. Our focus is, however, on infinitesimal external stress following the definition of the shear viscosity of the equilibrium system as given by linear response in that limit.

\bibitem{sausset}
F.~Sausset, G.~Biroli, and J.~Kurchan, ``Do solids flow?,'' {\em Journal of Statistical Physics}, vol.~140, p.~718, 2010.

\bibitem{Note12}
The full many body probability density matrix evolves as \begin {eqnarray} \rho _{\Lambda } (t') = e^{-i {\protect \hat {\protect \cal {L}}} t'} \rho _{\Lambda }(0). \end {eqnarray} Classically, the Liouvillian is given by \begin {eqnarray} {\protect \hat {\protect \cal {L}}} \cdot = i \{ H, \cdot \}_{P.B.} \end {eqnarray} with $\{ , \}_{P.B.}$ denoting a Poisson bracket. Quantum mechanically, \begin {eqnarray} {\protect \hat {\protect \cal {L}}} \cdot = \protect \frac {1}{\hbar } [H, \cdot ]. \end {eqnarray}.

\bibitem{NZ1}
S.~Nakajima, ``On quantum theory of transport phenomena: Steady diffusion,'' {\em Progress of Theoretical Physics}, vol.~20, p.~948, 1958.

\bibitem{NZ2}
R.~Zwanzig, ``Ensemble method in the theory of irreversibility,'' vol.~33, p.~1338, 1960.

\bibitem{Lindblad}
G.~Lindblad, ``On the generators of quantum dynamical semigroups,'' {\em Commun. Math. Phys.}, vol.~119, p.~48, 1976.

\bibitem{Bello}
A.~Bello, E.~Laredo, and M.~Grimau {\em Phys. Rev. B}, vol.~60, p.~12764, 1999.

\bibitem{Nick-to-be}
N.~B. Weingartner~et al. {\em In preparation}.

\bibitem{Keys}
A.~S. Keys, J.~P. Garrahan, and D.~Chandler, ``Calorimetric glass transition explained by hierarchical dynamic facilitation,'' {\em Proceedings of the National Academy of Sciences}, vol.~110, p.~4482, 2013.

\bibitem{Richert}
R.~Richert, K.~Duvvuri, and L.-T. Duong, ``Dynamics of glass-forming liquids. vii. dielectric relaxation of supercooled tris-naphthylbenzene, squalane, and decahydroisoquinoline,'' {\em J. Chem. Phys.}, vol.~118, p.~1828, 2003.

\bibitem{liquid-liquid}
H.~Tanaka, ``Liquid-liquid transitions and polyamorphism,'' {\em J. Chem. Phys.}, vol.~153, p.~130901, 2020.

\bibitem{Jing}
J.~Xue~et al. {\em In preparation}.

\bibitem{Ryan1}
R.~Soklaski, Z.~Nussinov, Z.~Markow, K.~F. Kelton, and L.~Yang, ``Connectivity of icosahedral network and a dramatically growing static length scale in cu-zr binary metallic glasses,'' {\em Physical Review B}, vol.~87, p.~184203, 2013.

\bibitem{shear1}
N.~B. Weingartner, R.~Soklaski, K.~F. Kelton, and Z.~Nussinov, ``Dramatically growing shear rigidity length scale in the supercooled glass former nizr2,'' {\em Physical Review B}, vol.~93, p.~214201, 2016.

\bibitem{shear2}
N.~B. Weingartner and Z.~Nussinov, ``Probing local structure in glass by the application of shear,'' {\em J. Stat. Mech}, vol.~2016, p.~0944001, 2016.

\bibitem{BO}
P.~Steinhardt, D.~R. Nelson, and M.~Ronchetti, ``Bond-orientational order in liquids and glasses,'' {\em Physical Review B}, vol.~28, p.~784, 1983.

\bibitem{structure_ML}
P.~Ronhovde, S.~Chakrabarty, M.~Sahu, K.~F. Kelton, N.~A. Mauro, K.~K. Sahu, and Z.~Nussinov, ``Detecting hidden spatial and spatio-temporal structures in glasses and complex physical systems by multiresolution network clustering,'' {\em The European Physics Journal E}, vol.~34, p.~105, 2011.

\bibitem{MBL}
D.~Basko, I.~Aleiner, and B.~Altshuler, ``Metal-insulator transition in a weakly interacting many-electron system with localized single-particle states,'' {\em Annals of Physics}, vol.~321, p.~1126, 2006.

\bibitem{MBL_rev}
D.~A. Abanin, E.~Altman, I.~Bloch, and M.~Serbyn, ``Colloquium: Many-body localization, thermalization, and entanglement,'' {\em Rev. Mod. Phys.}, vol.~91, p.~021001, 2019.

\bibitem{scar}
E.~J. Heller, ``Bound-state eigenfunctions of classically chaotic hamiltonian systems: Scars of periodic orbits,'' {\em Physical Review Letters}, vol.~53, p.~1515, 1984.

\bibitem{scar_rev}
A.~Chandran, T.~Iadecola, V.~Khemani, and R.~Moessner, ``Quantum many-body scars: A quasiparticle perspective,'' {\em Annual Review of Condensed Matter Physics}, vol.~14, p.~443, 2023.

\bibitem{Note13}
In Refs. \cite {1parameter,PS,longrange,critical,local_structure}, it was assumed that the {\protect \it full many body} system remains in the supercooled state for divergent times. In actual experiments, the supercooled system exchanges heat with its surroundings and transitions into the crystalline equilibrium state at times $t_{\protect \sf xtal}$ set by the driving free energy for nucleation and growth \cite {KG}. If heat could not be exchanged with the environment then any nucleation might heat up the system- thus in turn slowing crystallization and possibly remelting the sample. In such an idealized setting, the supercooled system might avoid crystallization for a very long time (potentially even at divergent times as implicitly assumed in \cite {1parameter,PS,longrange,critical,local_structure} that considered ideal perfectly isolated metastable supercooled liquids to remain unchanged). In the current work, we focused, by contrast, on the empirically pertinent near-stationarity of few body observables.
  Local uncorrelated noise from the environment may (and does) trigger nucleation and bona-fide equilibriation to the crystalline state (at sufficiently long times) and can thus inhibit long-range correlations.

\bibitem{Note14}
The astute reader will indeed recognize that given Eq. (\ref {r2w}), this form of $w_{ij}$ implies that the connected two-body correlations indeed decay with the distance with a correlation length $\xi $.

\bibitem{BBGKY1}
N.~N. Bogoliubov, {\em The dynamical theory in statistical physics in Studies in Statistical Mechanics Vol. 1, Edited by J. de Boer and G. E. Uhlenbeck}.
\newblock North Holland, Amsterdam, 1962.

\bibitem{BBGKY2}
M.~Born and H.~Green, {\em A General Kinetic Theory of Liquids}.
\newblock Cambridge University Press, 1949.

\bibitem{BBGKY3}
J.~K. Kirkwood, ``The statistical mechanical theory of transport processes i. general theory,'' {\em J. Chem. Phys.}, vol.~14, p.~180, 1946.

\bibitem{BBGKY4}
J.~K. Kirkwood, ``The statistical mechanical theory of transport processes ii. transport in gases,'' {\em J. Chem. Phys.}, vol.~15, p.~72, 1947.

\bibitem{BBGKY5}
J.~Yvon, {\em La Théorie des Fluids et L’équation D’etat: Actualités Scientificues et Industrielles}.
\newblock Hermann and Cie, Paris, 1935.

\bibitem{LR1}
B.~Nachtergaele and R.~Sims, ``Lieb-robinson bounds and the exponential clustering theorem,'' {\em Communications in Mathematical Physics}, vol.~265, p.~119, 2006.

\bibitem{LR}
E.~H. Lieb and D.~W. Robinson, ``The finite group velocity of quantum spin systems,'' {\em Communications in Mathematical Physics}, vol.~28, p.~251, 1972.

\bibitem{kirkwood1}
J.~G. Kirkwood, ``Statistical mechanics of fluid mixtures,'' {\em J. Chem. Phys.}, vol.~3, p.~300, 1935.

\bibitem{kirkwood2}
J.~G. Kirkwood and E.~M. Boggs, ``The radial distribution function in liquids,'' {\em J. Chem. Phys.}, vol.~10, p.~394, 1942.

\bibitem{Singera}
A.~Singera, ``Maximum entropy formulation of the kirkwood superposition approximation,'' {\em J. Chem. Phys.}, vol.~121, p.~3657, 2004.

\bibitem{Note15}
In the simplest (Kirkwood) approximation, the associated correlation function of three particles is equal to the product of the three pair correlation functions, $g({\protect \bf {x}}_1,{\protect \bf {x}}_2,{\protect \bf {x}}_3) = g({\protect \bf {x}}_1,{\protect \bf {x}}_2)g({\protect \bf {x}}_2,{\protect \bf {x}}_3)g({\protect \bf {x}}_1,{\protect \bf {x}}_3)$. Such an approximation neglects subtle higher body correlations.

\bibitem{LL1}
L.~D. Landau and E.~M. Lifshitz, {\em Mechanics, Course of Theoretical Physics}, vol.~1.
\newblock Butterworth-Heinenann, 1976.

\end{thebibliography}

\end{document}